\documentclass[referee]{jfm}

\usepackage{graphicx}
\usepackage{newtxtext}
\usepackage{newtxmath}
\usepackage{natbib}
\usepackage{hyperref}
\usepackage{subcaption}
\usepackage{tabularx}
\usepackage{booktabs}
\usepackage{caption}
\usepackage{float}
\hypersetup{
    colorlinks = true,
    urlcolor   = blue,
    citecolor  = black,
}

\newcommand{\RomanNumeralCaps}[1]
\linenumbers

\title{On Deep-Learning-Based Closures for Algebraic Surrogate Models of Turbulent Flows}

\author{B. Eiximeno\aff{1,2}
  \corresp{\email{benet.eiximeno@bsc.es}},
  M. Sanchis-Agudo\aff{3}, A. Miró\aff{1,2}, I. Rodriguez\aff{2}, R. Vinuesa\aff{3} \and O. Lehmkuhl\aff{1}}

\affiliation{\aff{1}Barcelona Supercomputing Center, Spain
\aff{2}Universitat Politècnica de Catalunya, Spain \aff{3}FLOW, Engineering Mechanics, KTH Royal Institute of Technology, Sweden}

\begin{document}
\maketitle

\begin{abstract}
A deep-learning-based closure model to address energy loss in low-dimensional surrogate models based on proper-orthogonal-decomposition (POD) modes is introduced. Using a transformer-encoder block with easy-attention mechanism, the model predicts the spatial probability density function of fluctuations not captured by the truncated POD modes. The methodology is demonstrated on the wake of the Windsor body at yaw angles of $\delta = [2.5^\circ,5^\circ,7.5^\circ,10^\circ,12.5^\circ]$, with $\delta = 7.5^\circ$ as a test case. Key coherent modes are identified by clustering them based on dominant frequency dynamics using Hotelling’s $T^2$ on the spectral properties of temporal coefficients. These coherent modes account for nearly 60\% of the total energy while comprising less than 10\% of all modes. A common POD basis is created by concatenating coherent modes from training angles and orthonormalizing the set, reducing the basis vectors from 142 to 90 without losing information. Transformers with different size on the attention layer, (64, 128 and 256), are trained to model the missing fluctuations. Larger attention sizes always improve predictions for the training set, but the transformer with an attention layer of size 256 overshoots the fluctuations predictions in the test set because they have lower intensity than in the training cases. Adding the predicted fluctuations closes the energy gap between the reconstruction and the original flow field, improving predictions for energy, root-mean-square velocity fluctuations, and instantaneous flow fields. The deepest architecture reduces mean energy error from 37\% to 12\% and decreases the Kullback--Leibler divergence of velocity distributions from ${\cal{{D_{KL}}}}=0.2$ to below ${\cal{{D_{KL}}}}=0.026$.
\end{abstract}



\section{Introduction}
Surrogate models are data-driven computational techniques used in various scientific and engineering fields to approximate complex systems or functions. These models serve as simpler substitutes for both experiments and computationally expensive simulations, thus providing quicker, yet sufficiently accurate results \citep{GangSun}. Surrogate models are mainly utilized to estimate the optimum-product solution or as instrumental tools to evaluate the performance in the initial stages of the vehicle development because they reduce the resource requirements for design exploration \citep{yuichi2011, yondo_review_2018}.

In the particular case of fluid-dynamics applications, surrogates are tipically built on a reduced space due to the complexity and high dimensionality of the original phenomenon \citep{yondo_review_2018}. The dimensionality reduction can be done either with algebraic methods, {\it{e.g.}} the proper-orthogonal decomposition (POD) \citep{lumley_rational_2004}, or employing deep-learning-based techniques. POD was first introduced in fluid dynamics by \cite{lumley_rational_2004} to express the chaotic turbulent motions into modes representing some portion of the total fluctuating energy of the flow. \cite{sirovich} explored the relationship between POD and the dominant features of the flow, and showed that POD is a relevant tool for the study of vortex dynamics in all types of fluid flows. Recently, other modal decompositions have been introduced in order to obtain modes that are associated with a single frequency instead of the range of frequencies present in the time series of the temporal coefficients in POD. Among these new techniques, the most popular are dynamic-mode decomposition (DMD) \citep{schmid2010dynamic} and spectral proper-orthogonal decomposition (SPOD) \citep{towne2018spectral}. Note that while POD and SPOD rank the modes in terms of their contribution to the reconstruction of the original flow, DMD obtains modes classified in terms of their dynamical importance to minimize errors in the reconstruction.

Alternatively, deep-learning methods for dimensionality reduction are based on unsupervised-learning methodologies such as autoencoders. There are application examples of several autoencoder architectures for dimensionality reduction in fluid dynamics, including vanilla \citep{Eivazi2020, Murata2019}, hierachical \citep{Fukami2020}, physics-assimilated \citep{Zhang2023} and variational autoencoders \citep{Eivazi2022Orthogonal, WANG2024109254,solera2024beta,Akkari2022}. All of them are able to capture the non-linear behaviour of dynamical systems with a higher compression capacity than any POD-based methodology thanks to the excellent capabilities of spatial convolutions for non-linear feature extraction \citep{brunton_2020,vinuesa2022enhancing}. 

It is particularly relevant to mention that $\beta$-variational autoencoders based on convolutional neural networks (CNN-$\beta$VAEs) have been used successfully to obtain a disentangled latent representation of turbulent fluid flows. For instance, \citet{Eivazi2022Orthogonal} compressed the turbulent flow around a simplified urban environment into 5 orthogonal latent variables containing more than the 85\% of the flow energy. However, the need of convolutional layers restricts the usage of this technique to geometries that can be represented on a regular grid. On the other hand, algebraic decompositions can be used on unstructured grids at the cost of losing a significant amount of the energy of the system. A good illustration of this is the aforementioned study from \citet{Eivazi2022Orthogonal}, where 5 POD modes barely recover 30\% of the flow energy. Accurately capturing all the fluctuations in a turbulent flow would require selecting nearly all the modes of the system. 

\citet{COUPLET_SAGAUT_BASDEVANT_2003} proved that large-index POD modes drain energy from the more significant modes, yielding an energy-cascade structure. Such a modal-energy redistribution suggests that reduced-order models (ROMs) can be built on a small number of significant modes that represent the majority of flow features and the contribution of the rest of modes can be modelled as an additional term to the ROM. This conclusion has led to an intense research on closures for reduced-order models based on Galerkin and Petrov--Galerkin projections of the Navier--Stokes equations. These models constitute a fundamental pillar for the stability of the projection \citep{STABILE2018273,kaptanoglu} and have been traditionally inspired by sub-grid scale models such as the ones used in large-eddy simulations (LESs) \citep{WANG201210,HIJAZI2020109513, imtiaz2020nonlinear}. More recently, such closures have been modelled with data-driven techniques such as probabilistic neural networks \citep{Maulik}. A recent review and comparison of data-driven methods for ROM closures can be found in \citet{PRAKASH2024116930}. 

The main goal of this manuscript is to present a new data-driven model capable of recovering the energy loss due to modal truncation in POD. Instead of working in the reduced space as the aforementioned closures, this work is focused on learning the spatial probability density function of the difference between the original field and the POD reconstruction using only the most significant modes with a transformer model \citep{vaswani2017attention}. A transformer is a deep-neural-network architecture initially developed in the field of natural-language processing (NLP). Since then, it has revolutionized many areas of machine learning thanks to its attention mechanism, which enables identifying long-range dependencies in the data more effectively than traditional models \citep{Yousif_Zhang_Yu_Vinuesa_Lim_2023}. The rationale behind the approach proposed is to build a reduced-order model capable of predicting the most significant features of the flow, which are fully dependent on the geometry and initial conditions, and then add a separate correction for the smaller turbulent scales. The methodology is tested on the turbulent wake of the flow past the Windsor body \citep{Littlewood2010}, which is a simplified square-back vehicle. The model is designed to produce a closure valid for any free-stream-velocity direction in a yaw-angle range $2.5^{\circ} \leq \delta \leq 12.5^{\circ}$. The objective of the closure is to improve the POD reconstruction of the root-mean-square values of the stream-wise velocity fluctuations. This test case is highly relevant for the automotive industry because in any road vehicle the drag force increases linearly for yaw angles in the range of $0^{\circ} \leq \delta \leq 15^{\circ}$ \citep{Howell2015}. This drag increase is completely independent of the zero-yaw drag, thereby making it impossible to extrapolate the performance in cross-flow conditions from the parallel-flow case \citep{Howell2015}. Hence, car manufacturers need to evaluate the aerodynamic performance under yawed flows in the development loop of a new vehicle \citep{DHooge2014}. The development could be massively accelerated by using a surrogate model instead of re-running the simulations and wind-tunnel tests that are needed to characterize the aerodynamic performance of a road vehicle \citep{Zhang2006} at every angle of interest.

There are a number of studies in the literature that propose models able to evaluate dependence of the forces and moments on the yaw angle. For instance, \citet{gong2012surrogate} built a surrogate model based on the Kriging interpolation technique, which obtains the optimal wind-deflector geometry in a tractor trailer to reduce drag in crosswind situations. Similarly, \citet{GHOREYSHI2014222} developed a model able to predict the dependence of the forces and moments on the Mach number, the angle of attack and the side-slip angle of an aircraft, while \citet{Zhang2021} presented a model to predict the derailment coefficient of a train in crossflows. Lately, \citet{eiximeno2024toward} developed a model to interpolate the mean base pressure in the Windsor body. To the authors' knowledge, there are no models that can evaluate changes in high order statistics on unseen flow conditions, hence, there are no models able to predict the velocity fluctuations with the yaw angle.

The rest of this manuscript is organized as follows: \autoref{sec:methodology} describes how the closure is formulated, how the significant POD modes are selected and how the model is extended to multiple flow conditions; then, \autoref{sec:results} shows the accuracy of the closure in the wake behind the Windsor body; and finally, \autoref{sec:conclusions} summarizes the main findings of the manuscript.

\section{Methodology}
\label{sec:methodology}
This section describes the methods used in the manuscript, including the Windsor-body dataset employed to test the methodology, a mathematical definition of POD and the selection of the most significant modes, together with an explanation of the model used to add the energy from the truncated modes.

\subsection{Dataset description}
\begin{figure}
    \centering
    \includegraphics[width=\textwidth]{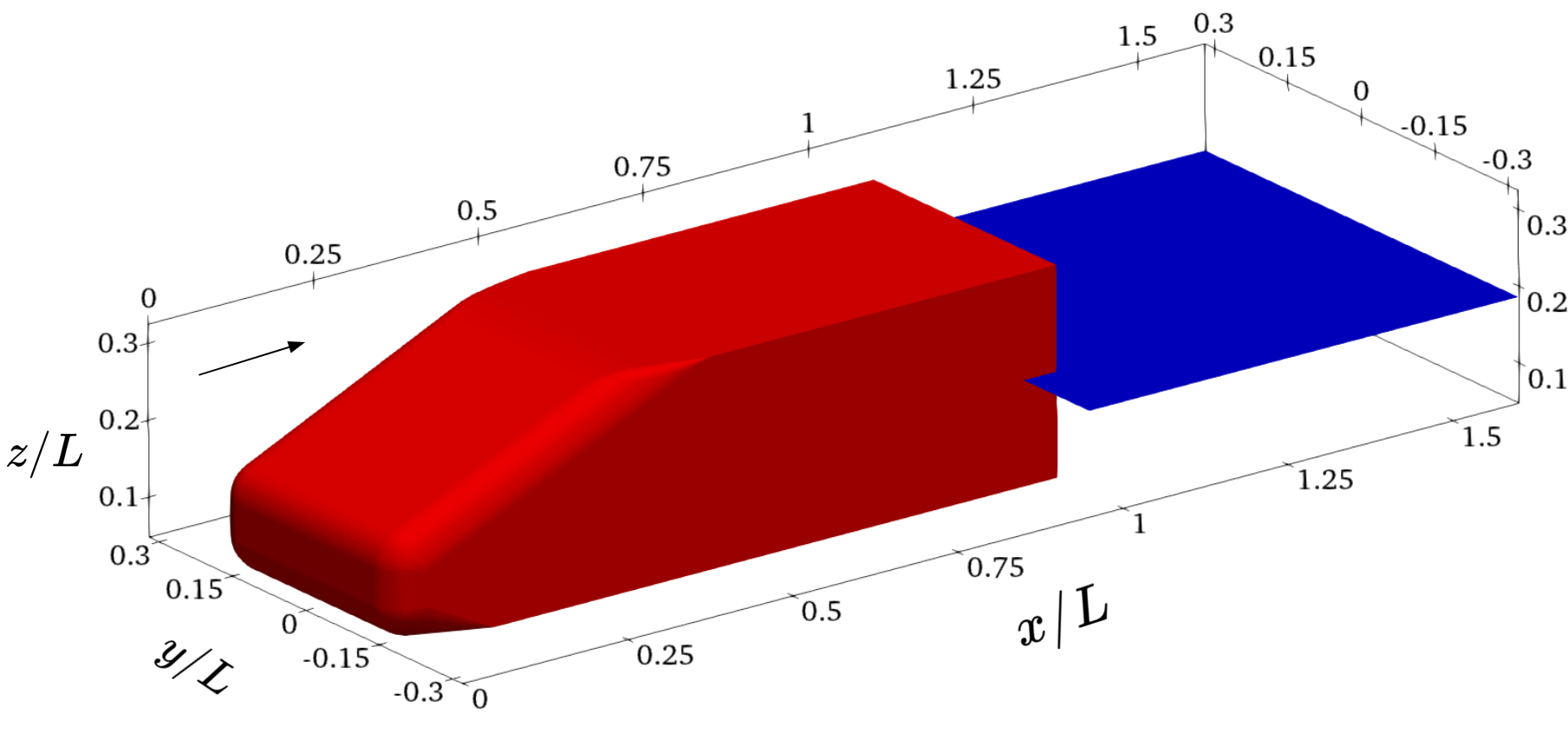}
    \caption{(Red) geometry of the Windsor body and (blue) working plane where the data is interpolated to develop the model. The plane is perpendicular to the vertical axis and is located at $z/L=0.186$. The arrow indicates the flow direction.}\label{fig:geom}
\end{figure}

The test dataset is the turbulent wake behind the Windsor body, the simplified square-back vehicle depicted in \autoref{fig:geom}, at a Reynolds number of $Re_L = U_{\infty}L/\nu = 2.9\times10^6$, where $U_{\infty}$ is the magnitude of the free-stream velocity, $L$ is the length of the model and $\nu$ is the kinematic viscosity of the fluid. The data was generated by means of wall-modeled large-eddy simulations at yaw angles of $\delta = [2.5^{\circ}, 5^{\circ}, 7.5^{\circ}, 10^{\circ} \text{ and } 12.5^{\circ}]$. For the simulations, the spatially filtered incompressible Navier--Stokes equations, 

\begin{align}
    \label{eq:continuity} \frac{\partial\overline {u}_i}{\partial x_i} &= 0,  \\
    \label{eq:momentum} \frac{\partial \overline{u}_i} {\partial t} +
    \frac{\partial \overline{u}_i \overline{u}_j}{\partial x_j}
    - \nu \frac{\partial^2 \overline{u}_i}{\partial x_j \partial x_j}
    + \rho^{-1}   \frac{\partial \overline{p}}{\partial x_i}  &=
    - \frac{\partial {\cal T}_{ij}}{\partial x_j},
\end{align}

were numerically integrated using SOD2D (Spectral high-Order coDe 2 solve partial Differential equations) \citep{Silva2023Sod2d}, a low-dissipation spectral-element-method (SEM) code \citep{Gasparino_Muela_Lehmkuhl}.

In the equations above $x_i$ are the spatial coordinates (or $x$, $y$, and $z$),  $ {u}_i$ (or ${u}$, ${v}$, and ${w}$) stands for the velocity components and  ${p}$  is  the pressure. Note that $\rho$ is the density of the fluid. The filtered variables are represented by $(\overline {\cdot})$.
The right-hand-side term in \autoref{eq:momentum} represents the sub-grid stresses, and its  anisotropic part is expressed as,
  
\begin{equation} \label{sgss}
    {\cal T}_{ij}- \frac{1}{3}{\cal T}_{kk} \delta_{ij} =
    -2\nu_{sgs} \overline{{\cal S}}_{ij}, 
 \end{equation}
where the large-scale rate-of-strain tensor $\overline{{\cal S}}_{ij}$ is evaluated as $\overline{{\cal S}}_{ij}=\frac{1}{2} \left (g_{ij}+g_{ji}\right )$, with $g_{ij}=\partial \overline{u}_{i}/\partial x_j$ and  $\delta_{ij}$ being the  Kronecker delta. Here, the unresolved scales are modelled using the local formulation of the integral length-scale approximation (ILSA) \citep{LEHMKUHL2019108422}. The near wall region was modelled using the Reichardt wall-law \citep{reichardt1951vollstandige} with an exchange location in the 5th node \citep{lehmkuhl2018large}.

After the initial transients had been washed out, all simulations were run for 60 additional convective time units, $t=60L/U_{\infty}$, to collect 660 snapshots. The data for the model assessment was interpolated into the plane represented in \autoref{fig:geom}. This plane is perpendicular to the vertical axis, therefore it contains the dynamics of both the leeward and windward sides of the wake. It is located at $z/L=0.186$, which is half of the vehicle height when measured from the bottom of the body.

In the present work only a brief comparison of the fluid flow at the different yaw angles is shown to illustrate the different conditions in which the closure needs to be valid. For more details on the numerical model, grid and simulations accuracy, the reader is referred to the previous work by \citet{eiximeno2024toward} on the development of a surrogate model for the base pressure of the Windsor body. 

\begin{figure}
    \centering
    \includegraphics[width=\textwidth]{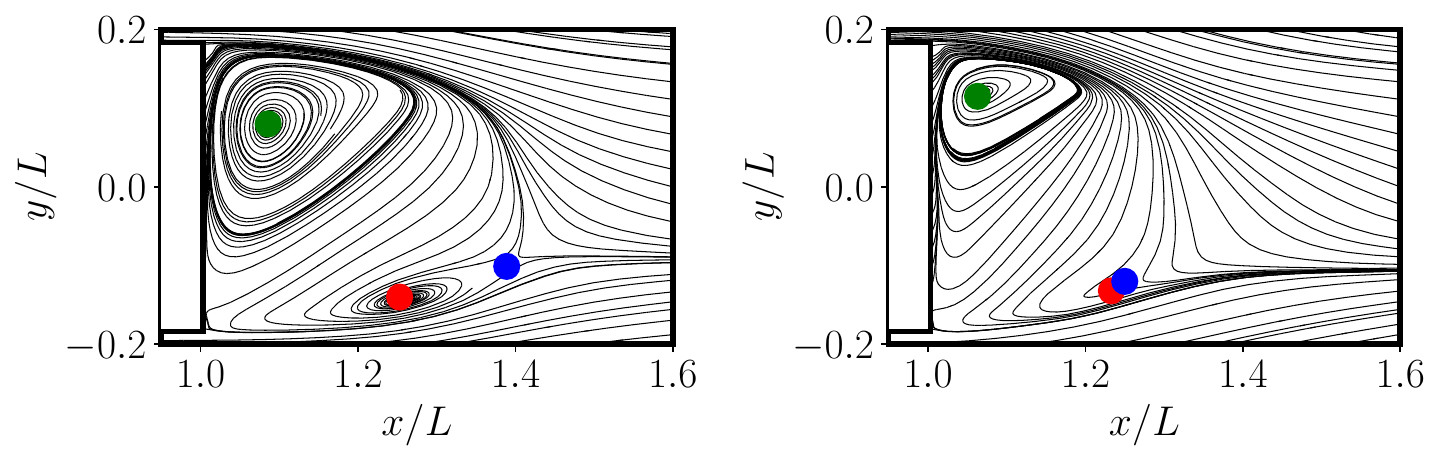} 
    \caption{Streamline comparison between $\delta=2.5^{\circ}$ (left) and $\delta=12.5^{\circ}$ (right) at the plane $z/L=0.186$. The green, red and blue dots represent the core of the leeward side vortex, the core of the windward side vortex and the saddle point, respectively.}
    \label{fig:vortexs}
\end{figure}

In terms of the averaged flow, the wake of square-back bluff bodies in a yawed free stream flow is dominated by two vortices: one on the leeward side ($y/L>0$) and one on the windward side ($y/L<0$), as it was shown by \cite{BOOYSEN2022110562}. In \autoref{fig:vortexs}, the flow streamlines are plotted for $\delta=2.5^{\circ}$ and $\delta=12.5^{\circ}$. As reported by \citet{BOOYSEN2022110562}, the vortex on the leeward side dominates the recirculation and gains intensity over the windward vortex as the yaw angle increases. This effect moves the vortex centers and the saddle point to the leeward side of the vehicle and closer to the body.

\begin{figure}
    \centering
    \includegraphics[width=\textwidth]{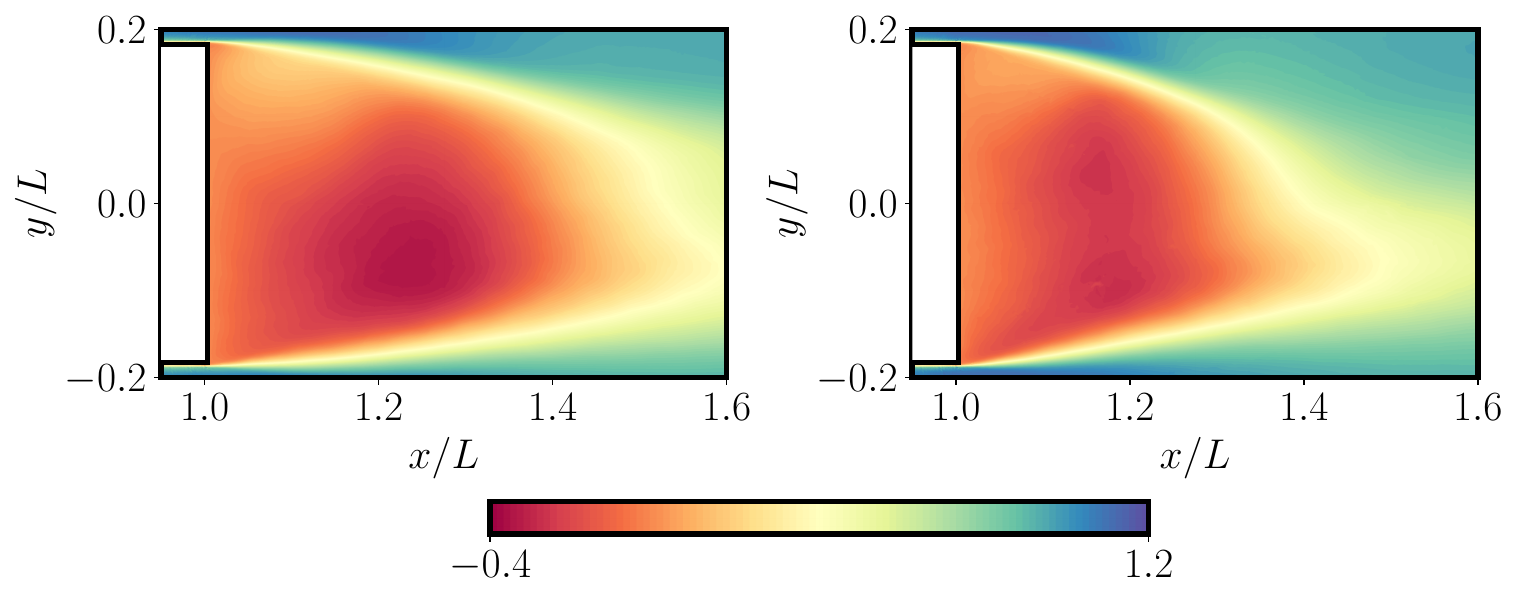}
    \caption{Mean streamwise velocity at $z/L=0.186$ for $\delta=2.5^{\circ}$ (left) and $\delta=12.5^{\circ}$ (right).}
    \label{fig:mean_veloc_contour}
\end{figure}

The changes in the vortex intensity have an effect on the recirculation length. \citet{LORITEDIEZ2020104145} identified, in a square-back Ahmed body, that this vortex interaction leads to a decrease on the recirculation length and, to a deflection of the recirculation bubble towards the leeward side of the wake. Similar to the square-back Ahmed body, in the Windsor body, this trend is also observed. This can be seen in \autoref{fig:mean_veloc_contour}, where the mean streamwise velocity $\overline{u}$ for the cases at $\delta=2.5^{\circ}$ and $\delta=12.5^{\circ}$ is plotted. Here the recirculation length varies from 0.41L to 0.28L.

\begin{figure}
    \centering
    \includegraphics[width=\textwidth]{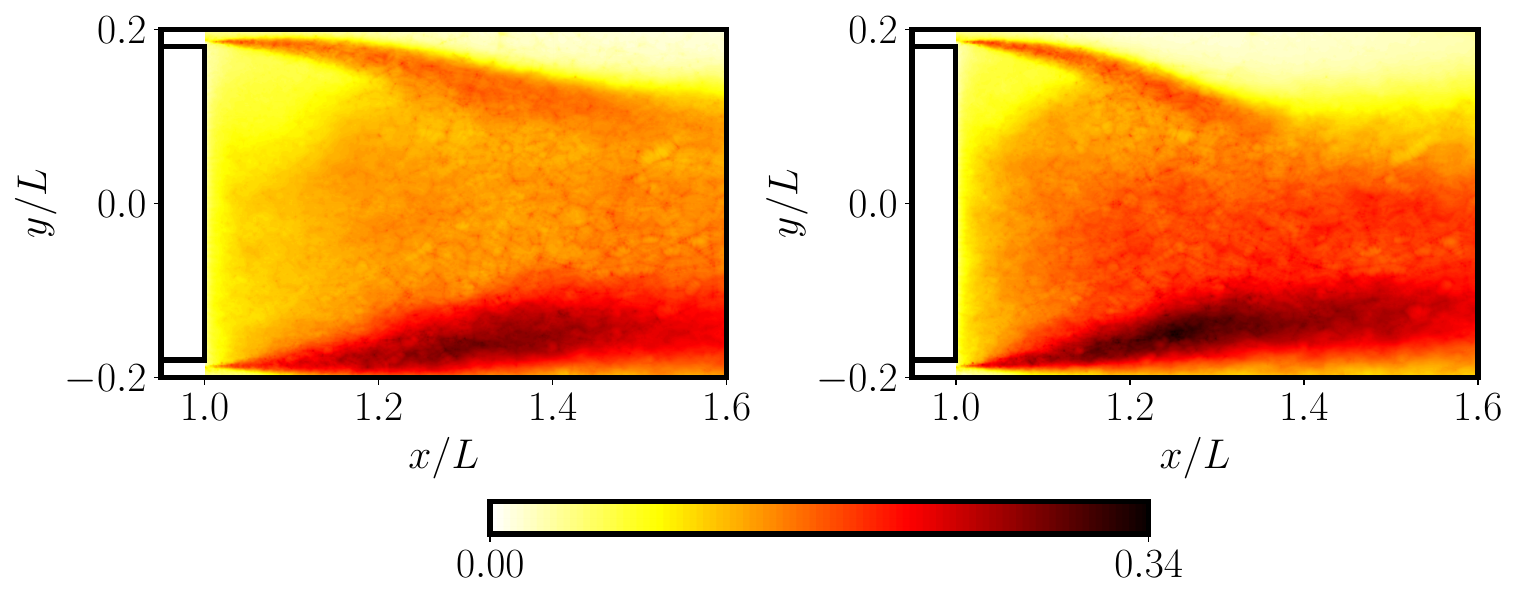}
    \caption{Root mean square of the streamwise velocity fluctuations at $z/L=0.186$ for $\delta=2.5^{\circ}$ (left) and $\delta=12.5^{\circ}$ (right).}
    \label{fig:rms_veloc_contour}
\end{figure}

In \autoref{fig:rms_veloc_contour} changes in the velocity fluctuations brought about with the yaw angle are illustrated by comparing the root-mean-square of the streamwise velocity fluctuations, $u_{\text{rms}}$ at both $\delta=2.5^{\circ}$ and $\delta=12.5^{\circ}$. \autoref{fig:rms_veloc_contour} shows that a larger yaw angle increases the entrainment of irrotational free-stream into the near wake, resulting in larger fluctuation intensity and a steeper shear layer angle on both sides of the vehicle. The latter leads to a narrower wake. This is in agreement with the findings from \citet{li2019} on a square-back Ahmed body.

\begin{figure}
    \centering
    \subfloat[]{\includegraphics[width=0.5\textwidth]{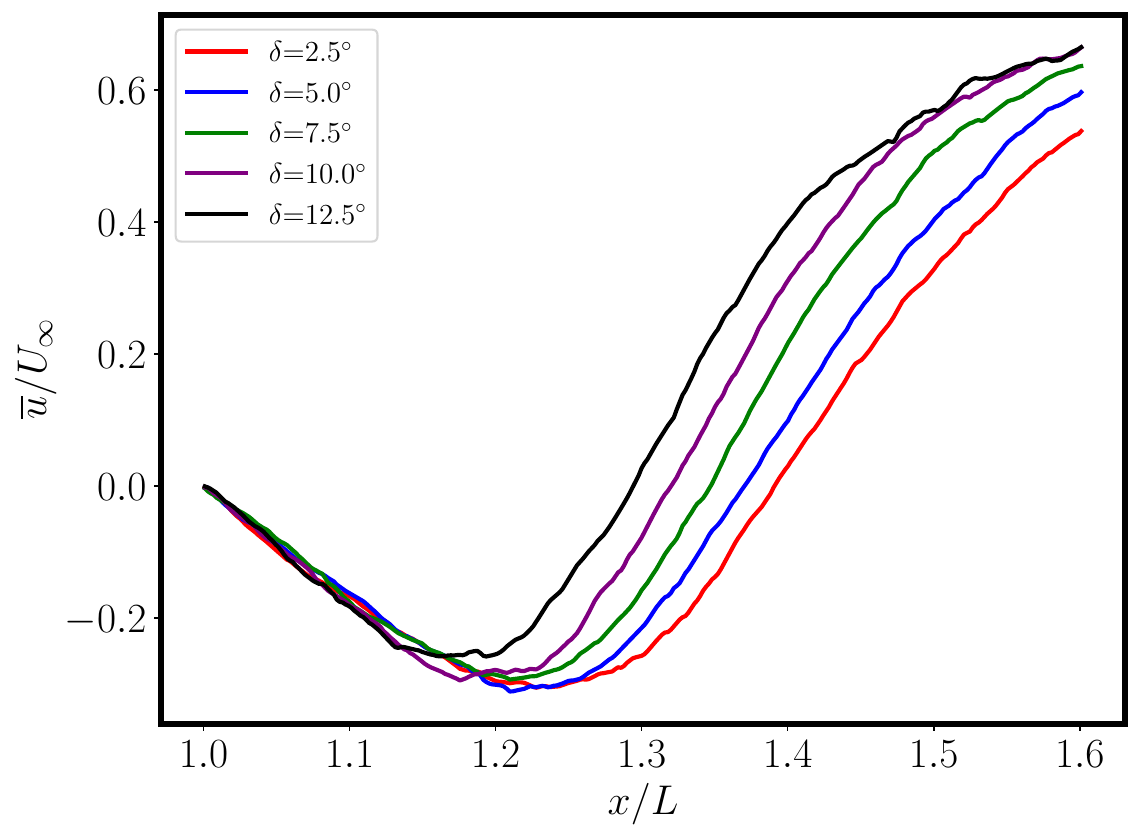}\label{fig:mean_veloc_all}}
    \subfloat[]{\includegraphics[width=0.5\textwidth]{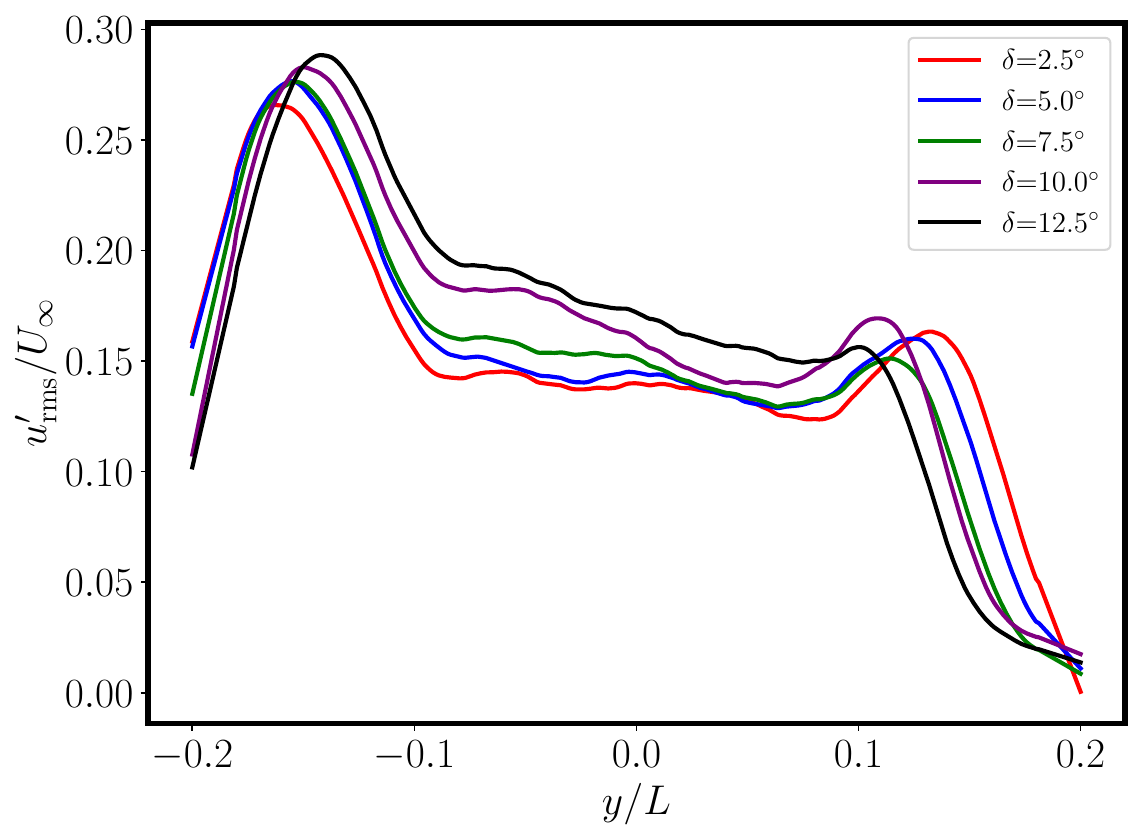}\label{fig:rms_veloc_all}}
   \caption{Mean streamwise velocity for all simulated angles at $y/L=0$ (a) and root-mean square value of the velocity fluctuations at $x/L=1.3$ (b).}\label{fig:veloc_all}
\end{figure}

The above changes in the mean flow with the yaw angle can also be observed when the mean streamwise velocity and its fluctuations are plotted along a streamwise line at $y/L=0$ and over a cross-stream line at $x/L=1.3$, respectively (see \autoref{fig:veloc_all}). For the objectives of the current work, it is relevant to remark that neither the fluctuations maxima nor their positions in the domain have a linear evolution with the yaw angle. Thus, it is not possible to derive a linear model to predict them.

\subsection{Proper-orthogonal decomposition (POD)}
POD is used in this work as a dimensionality-reduction technique. It is an efficient way to capture an infinite-dimensional process with a reduced number of modes \citep{holmes_low-dimensional_1997}. This method is based on finding a set of deterministic functions that characterize the dominant features of the system given by the field $\mathbf{F}(\mathbf{X}, t)$. This decomposition can be written as,
\begin{equation}
    \mathbf{F}(\mathbf{X}, t) = \sum_{i = 1}^{i = N}a_i(t) \mathbf{\Phi}_i(\mathbf{X}),
    \label{eqn:POD}
\end{equation}
where $N$ is the number of functions to decompose the field into. POD requires that the basis for the spatial modes is orthonormal, i.e., 
\begin{equation}
\int_{\mathbf{X}} \Phi_{i_1}(\mathbf{X}) \Phi_{i_2}(\mathbf{X}) \mathrm{d} x=\left\{\begin{array}{l}
1 \text { if } i_{1}=i_{2} \\
0 \text { otherwise }
\end{array}\right.
\end{equation}
and optimal, so that the the first $N_r$ vectors are the ones that reconstruct the database with the minimum possible error. 

In this work, the chosen method to perform POD is the singular-value decomposition (SVD). The SVD decomposes the initial snapshot matrix, $\mathbf{\cal{X}}$, into the left singular vectors, $\mathbf{\Psi}$, the singular values, $\mathbf{S}$, and the right singular vectors, $\mathbf{V}$,
\begin{equation}
    {\cal{X}} = \mathbf{\Psi} \mathbf{S}\mathbf{V}^{T}.
    \label{eqn:SVD}
\end{equation}
 
Each column of $\mathbf{\Psi}$ contains a spatial mode, $\mathbf{\Phi}_i(\mathbf{X})$ and each column of $\mathbf{V}$ gives the evolution of the time coefficient, $a_i(t)$, of the corresponding mode. The singular values are given in a diagonal matrix and are associated with the energy contribution of each mode in descending order. The higher the singular value, the more energy is contained in the mode. The POD analysis has been performed using pyLOM \citep{pyLOM}, a high-performance-computing reduced-order-modelling code that has a parallel and scalable algorithm for the singular-value decomposition \citep{eiximeno2024pylom}.

\subsection{On the significance of POD modes}

Turbulent flows are characterized by a flat-tail of singular values, making it difficult to set an energy threshold to select the modes onto which the data has to be projected. This threshold is set arbitrarily and is decided based on a trade-off between accuracy of the model and evaluation cost. To overcome this issue, in this work the selection of the relevant modes is based on their frequency content. The objective is to select only the modes that contain relevant information on the frequency of the coherent structures of the flow. This is achieved by identifying outlier modes in the power-spectral density (PSD) matrix of the temporal coefficients, $\mathbf{V}$. In other words, the selected modes are those that exhibit a frequency spectrum significantly different from the rest.

The PSD matrix of $\mathbf{V}$ is computed by performing the Lomb-Scargle periodogram to the temporal coefficient of each mode of the system. Then, the outlier modes are identified with principal-component analysis, PCA.  PCA is analogous to POD once the data has been normalized with its variance and centered to its mean. Since the PCA model may contain numerous components, its information is summarized using Hotelling's $T^2$:
\begin{equation}
\label{eqn:t2}
T^2=\sum_{a=1}^{a=A}\left(\frac{t_{i, a}}{s_a}\right)^2,
\end{equation}
where $t_{i,a}$ is the projection of the PSD of mode $i$ into the PCA component $a$ and $s_a$ is the covariance of that component. Note that $T^2$ can be seen as the distance from the center of the hyperplane formed by the components to the projection of the observation onto the hyperplane. The larger the $T^2$ value is, the more relevant frequency content the mode will have. Hence, the modes now can be selected with a $T^2$ threshold value that contains all the outliers.

\subsection{POD projection and reconstruction}
The POD basis for data projection is built using the spatial correlations of the $N_r$ modes corresponding to the frequency outliers. When working with $n$ different inlet conditions, one can find an optimal POD basis among them by concatenating the spatial correlations of the outlier modes from each case to create the following matrix $\mathbf{Y}$:
\begin{equation}
    \centering
    \label{eqn:y}
    \mathbf{Y} = \left[\mathbf{\Psi}_{0}\text{ }\mathbf{\Psi}_{1}\text{ }\dots\text{ }\mathbf{\Psi}_{n}\right].
\end{equation}
Then, POD is applied to matrix $\mathbf{Y}$ to find an orthonormal basis that contains the information of the selected modes for each of the inlet conditions:
\begin{equation}
    \centering
    \mathbf{Y} = \mathbf{\Psi}_Y \mathbf{S}_Y \mathbf{V}_Y^T.
\end{equation}
The resulting basis can be truncated as long as there are no information losses, i.e.\ the selected modes are able to recover more than 99\% of the energy.

The data matrix $\cal{X}$ can be projected now onto $\mathbf{U_Y}$ as follows:
\begin{equation}
    \centering
    \label{eqn:project}
    {\cal{\hat{X}}} = \mathbf{\Psi}_{Y}^{T} \cdot \cal{X},
\end{equation}
with the assurance that all coherent modes inside the inlet-conditions range are included in the reduced-order model. This operation reduces the dimensionality of the numerical data and sets a latent space for any surrogate-modelling applications. Such a surrogate model can be used to perform temporal predictions of the system or to evaluate its response to any condition in the evaluated range.

When a prediction, $\cal{\hat{X}_P}$, is reprojected back into the full-order space: 
\begin{equation}
    \centering
    \label{eqn:reproject}
    {\cal{X_P}} = \mathbf{\Psi}_{Y} \cdot \cal{\hat{X}_P},
\end{equation}
the main behavior of the system is captured, however, the model lacks the energy from the modes that were discarded during its construction. 

\subsection{Closure model}
The missing energy of the prediction arises from the error between the original data and the reconstruction from the truncated POD modes:
\begin{equation}
\label{eqn:err}
\centering
\cal{E} = \cal{X} - \cal{X}_{P}.
\end{equation}

To build a closure for the missing scales in the POD projection and reconstruction process, it is essential to understand the spatial and temporal distribution of this error. In other words, it is necessary to determine where and when this error is more likely to occur. The strategy followed in this work involves learning the evolution of the error as a function of the recovered fluctuations, $\cal{X_P}$, since this field contains all relevant information about the system's state at all points in the domain for the studied timestep. To achieve this, a transformer \citep{vaswani2017attention} encoder block is trained to minimize the difference between the actual error field, $\cal{E}$, and the predicted one, using the temporal series of $\cal{X_P}$ across all points in the domain. The training process employs a mean-squared-error loss function. Thus, if $\cal{X_P}$ is known for a given timestep, the transformer can predict the corresponding error field, $\cal{E}$. From now on, the error predicted by the transformer is represented as $\cal{E_T}$ and the error of the model after considering the closure is defined as:
\begin{equation}
    \centering
    \cal{E_M} = X - (X_P+E_T).
\end{equation}

The choice of using a transformer-based model is motivated by their ability to identify and predict the temporal dynamics of chaotic systems by capturing long-term dependencies in the data \citep{wupin,GENEVA2022272,sanchis2023easy}. Additionally, transformers are well-suited for forecasting time series based on other spatial variables \citep{wang2023convolution} through their variant known as visual transformers (ViT). Transformers can be seen as universal approximators to probability density functions \citep{furuya2024transformersuniversalincontextlearners}. Hence, the proposed model actually learns the joint probability density function (PDF) of $\cal{E}$ given $\cal{X}_{P}$, $p(\cal{E} \mid \cal{X}_{P})$. Furthermore, there exists an attention-only, Transformer \( T \) with attention normalization \( N \) such that, for any auto-regressive sequence \( (x_t)_{t \geq 1} \) converges exponentially fast as \( n \) goes to infinity, where \( n \) is the number of attention layers. Denoting
\[
\mathcal{E}(x_{1:t}) := \lim_{n \to +\infty} \mathcal{E}_n(x_{1:t}),
\]
one has
\[
\lim_{t \to +\infty} \left( \mathcal{E}(x_{1:t}) - x_{t+1} \right) = 0.
\]

For a more detailed study of the transformer's universality and the analytic intrinsics when approximating the theoretical measure the reader is referred to~\cite{geshkovski2024measuretomeasureinterpolationusingtransformers,sander2024understandinguniversalitytransformersnexttoken}.

The latter definition ensures that the system modeled by the transformer is statistically equivalent to the original one and that the closure will be generalizable as long as the joint PDF $p(\cal{E} \mid \cal{X}_{P})$ for a new set of data is similar to the original one. Such similarity is quantified using the Kullback--Leibler divergence, ${\cal{D_{KL}}}$:

\begin{equation}
    {\cal{D_{KL}}}(P_i || Q) = \int_{-\infty}^{\infty} P(x) \log\left(\frac{P_i(x)}{Q(x)}\right) dx,
\end{equation}
where $P_i(x)$ represents the joint PDF of the error for a single snapshot, while $Q(x)$ is the joint PDF for all snapshots included in the training.

In this study, the input signal has a time-delay dimension of 48 steps, which means that the input to the transformer is a sequence of 48 consecutive time steps of the POD reconstruction. The output is the prediction of the error of the first time instant of the input series. A time-space embedding module is added to each point time signal to incorporate temporal and spatial information before passing it to the transformer blocks, allowing the model to distinguish between the evolution of the velocity in different points at different time steps. An average pooling and a max-pooling layer are added to the time-space embedding. Both of them are one dimensional and have a stride of two steps.

\begin{figure}
    \centering
    \includegraphics[width=0.8\textwidth]{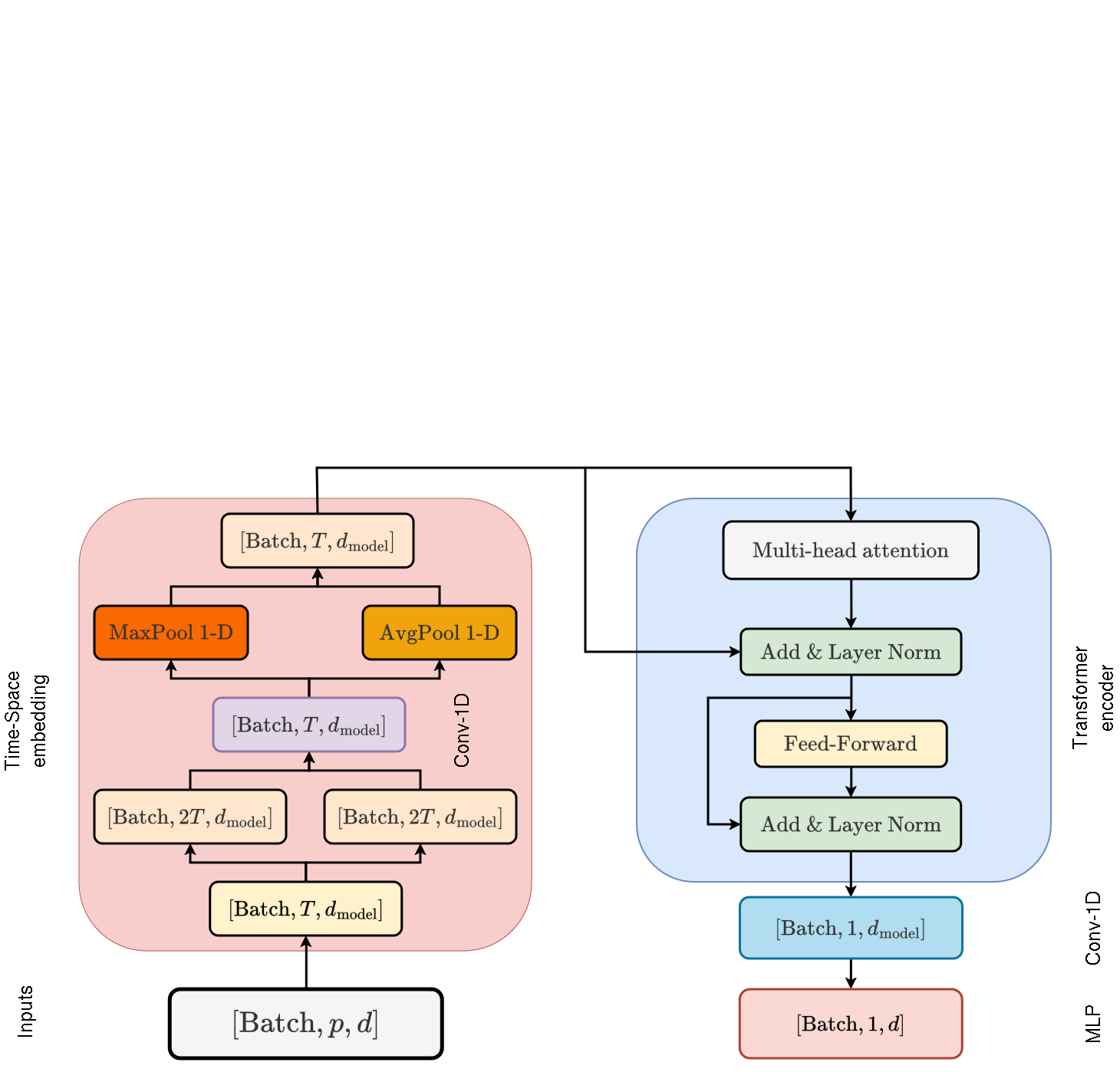}
    \caption{Easy-attention-based Transformer with time2space embedding. Figure adapted from \cite{sanchis2023easy}}
    \label{fig:trans}
\end{figure}

Three different architectures (\autoref{tab:architectures}) are tested to see the effect of the number of parameters on the closure accuracy. All of them are based on a single transformer encoder block (\autoref{fig:trans}) with eight attention heads followed by a feed-forward layer. The only change between the three architectures is the size of the attention layers. The shallowest architecture has an attention size of 64 and then increases to 128 and 256. Those layers are in charge of measuring the importance of different parts of the input sequence when making predictions \citep{bahdanau2014neural}. Note that, the dimension of the feed-forward layer is set to 128 in all three cases. This layer learns complex non-linear relationships between the input and output sequences. 

The choice of using multi-head attention is based on its oustanding performance over scaled dot product attention as it allows the model to jointly attend to information from different representation subspaces at different positions \citep{vaswani2017attention}. In particular, the current architecture employs the easy-attention mechanism \citep{sanchis2023easy}, which has demonstrated promising performance in predicting the temporal dynamics of chaotic systems, significantly outperforming the self-attention transformer \citep{solera2024beta}. The easy-attention mechanism originally presented by \cite{sanchis2023easy} is defined by the mapping $\mathbb{R}^{d_T \times d_S} \to \mathbb{R}^{d_T \times d_S}$, given by the equation $\mathbf{M} \rightarrow \hat{\mathbf{M}} = \alpha \mathbf{M} \mathbf{W_\text{V}}$, where both the pseudo-input, input after embedding, and output matrices have the same dimensions, the attention size (\autoref{tab:architectures}). In this formulation, $\alpha \in \mathbb{R}^{d_T \times d_T}$ and $\mathbf{W_\text{V}} \in \mathbb{R}^{d_S \times d_S}$ are matrices of trainable parameters, with $d_T$ representing the temporal feature dimension and $d_S$ the spatial feature dimension. Following the standard notation used in transformer architectures, $\mathbf{M} \cdot \mathbf{W_\text{V}}$ denotes the values, while the matrix $\alpha$ represents the attention weights. This mechanism, expressed as a kernel operation, can be formulated as:
\begin{equation} 
\hat{\mathbf{M}}(t,s) = \int_T \int_S \alpha(t,t') \mathbf{W_\text{V}}(s,s') \mathbf{M}(t',s')dt'ds'. 
\label{eq} 
\end{equation}
To extend the easy-attention mechanism to the multi-head attention strategy, we consider multiple attention heads so that each of them focus on different parts of the input space. In this case, the input \( \mathbf{M} \) is projected into multiple subspaces, allowing the model to attend to different sources of information simultaneously. For each attention head, we perform the same kernel operation as defined in the original mechanism, but with distinct sets of trainable parameters for the attention weights and value projections.

The multi-head version of the kernel operation can be written as:
\begin{equation}
    \hat{\mathbf{M}}^{(h)}(t,s) = \int_T \int_S \alpha^{(h)}(t,t') \mathbf{W_\text{V}}^{(h)}(s,s') \mathbf{M}(t',s') dt' ds',
\end{equation}
where \( h \in \{1, 2, \dots, H\} \) denotes the index of the attention head, \( H \) is the number of attention heads, and each \( \alpha^{(h)} \) and \( \mathbf{W_\text{V}}^{(h)} \) are distinct trainable parameters for the \( h \)-th attention head.
The final output of the multi-head attention is obtained by concatenating the outputs of each attention head:
\begin{equation}
    \hat{\mathbf{M}} = \text{Concat}(\hat{\mathbf{M}}^{(1)}, \hat{\mathbf{M}}^{(2)}, \dots, \hat{\mathbf{M}}^{(H)}).
\end{equation}

After the transformer block, a one-dimensional convolutional network of the same size as the attention layer and a fully connected layer of size of the number of points are added to decode the transformer output and form the final spatial prediction of the POD reconstruction error.

The training of each architecture was conducted along 3500 epochs, which required up to 8 hours and 15 minutes using an NVIDIA H100 GPU from the accelerated partition of the supercomputer MareNostrum 5 \citep{MareNostrum5LoginGuide}. An extensive discussion on the accuracy of each architecture is presented next in the results section.

\begin{table}
\centering
\begin{tabularx}{0.8\textwidth}{Xccc}
    \toprule
    \textbf{Parameter} & \textbf{Architecture 1} & \textbf{Architecture 2} & \textbf{Architecture 3} \\
    \midrule
    Input dimension        & 48         & 48         & 48         \\
    Output dimension       & 1          & 1          & 1          \\
    Time projection        & 128        & 128        & 128        \\
    Attention heads        & 8          & 8          & 8          \\
    Attention size         & 64         & 128        & 256        \\
    Feed forward layer     & 128        & 128        & 128        \\
    Activation function    & $\tanh$    & $\tanh$    & $\tanh$    \\
    Convolution layer      & 64         & 128        & 256        \\
    Fully connected layer  & 98304      & 98304      & 98304      \\
    Number of parameters   & 19,141,505 & 38,057,345 & 75,938,177 \\
    Size of the model (Mb) & 74         & 146        & 290        \\
    \bottomrule
\end{tabularx}
\caption{Summary of the three architectures considered in the present work}
\label{tab:architectures}
\end{table}

\section{Results}
\label{sec:results}
This section presents the performance of the designed closure model for the POD reconstruction of the turbulent wake behind the Windsor body. The dataset described in \autoref{sec:methodology} is split between the training, $\delta = [2.5^{\circ}, 5^{\circ}, 10^{\circ} \text{ and } 12.5^{\circ}]$, and test, $\delta = 7.5^{\circ}$, sets. The training set is used to build a common POD basis along the yaw-angle range and to train the transformer which predicts the reconstruction error. Then, the high-fidelity results at $\delta = 7.5^{\circ}$ are projected into that POD basis and reconstructed with the additional closure term from the transformer to assess its performance on unseen data. All results are obtained for the streamwise velocity fluctuations, therefore, $\cal{X}$ in \autoref{eqn:SVD} is equivalent to $u'$.

\subsection{POD common basis}
\begin{figure}
    \centering
    \subfloat[]{\includegraphics[width=0.49\textwidth]{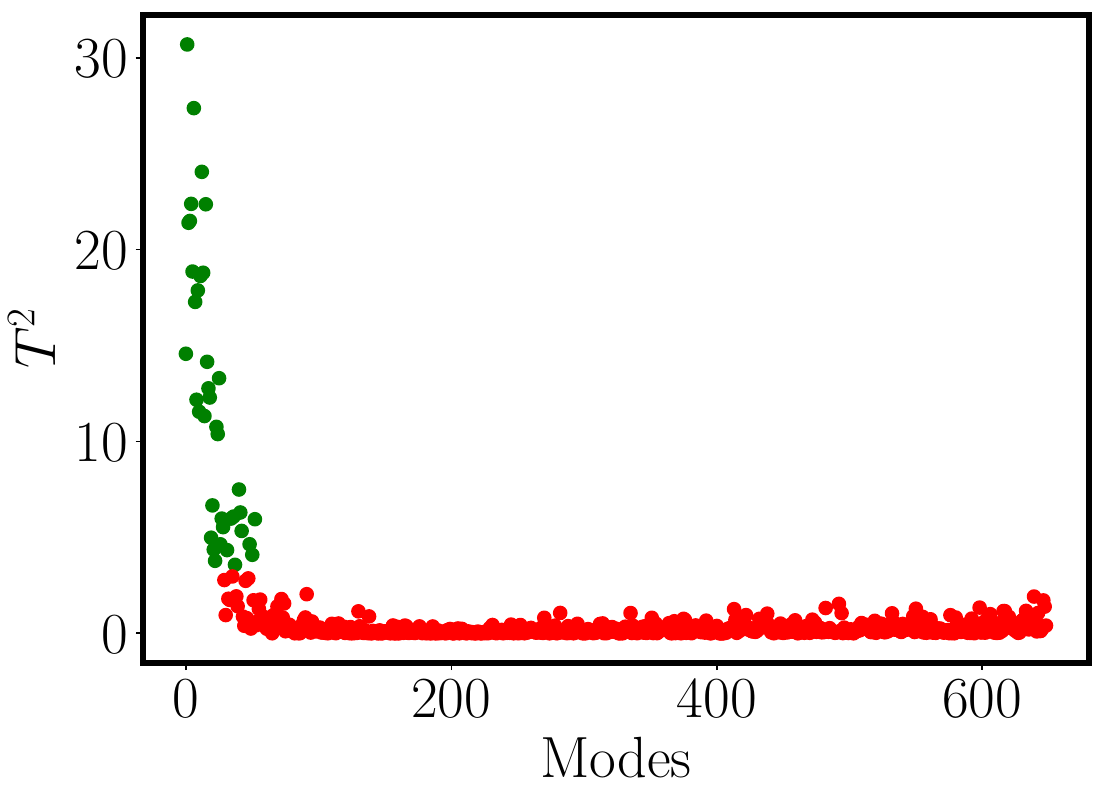}}
    \subfloat[]{\includegraphics[width=0.49\textwidth]{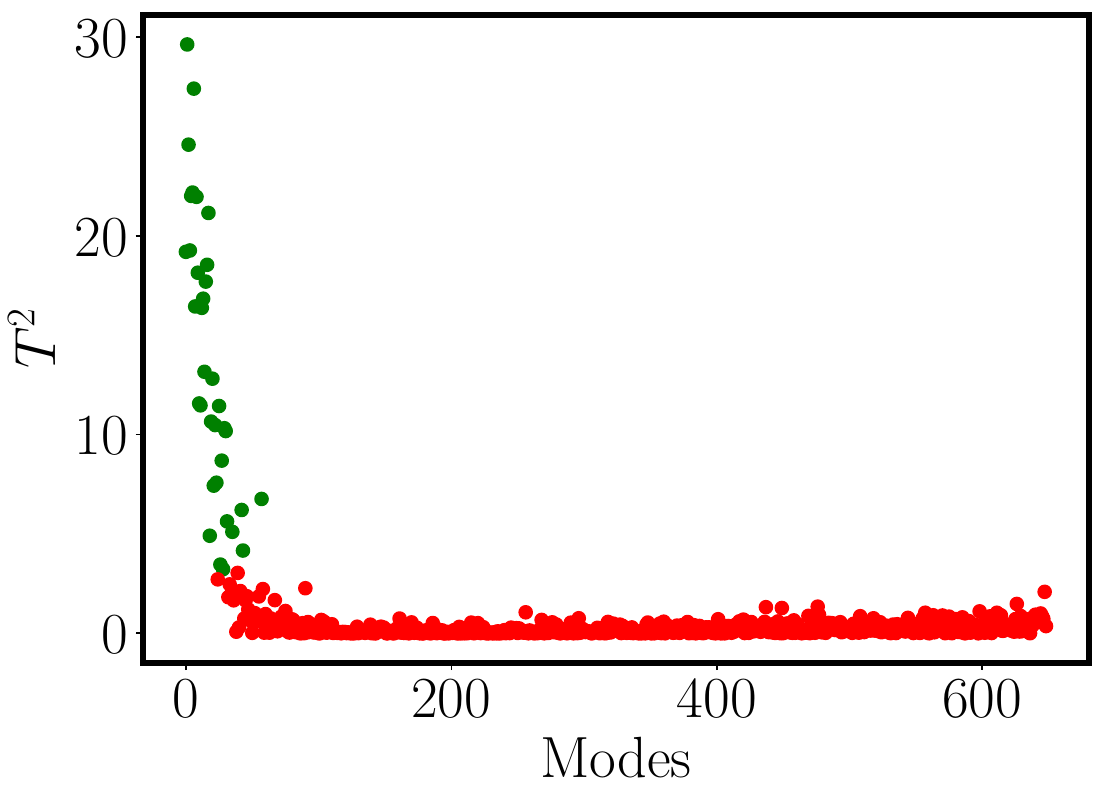}}\\
    \subfloat[]{\includegraphics[width=0.49\textwidth]{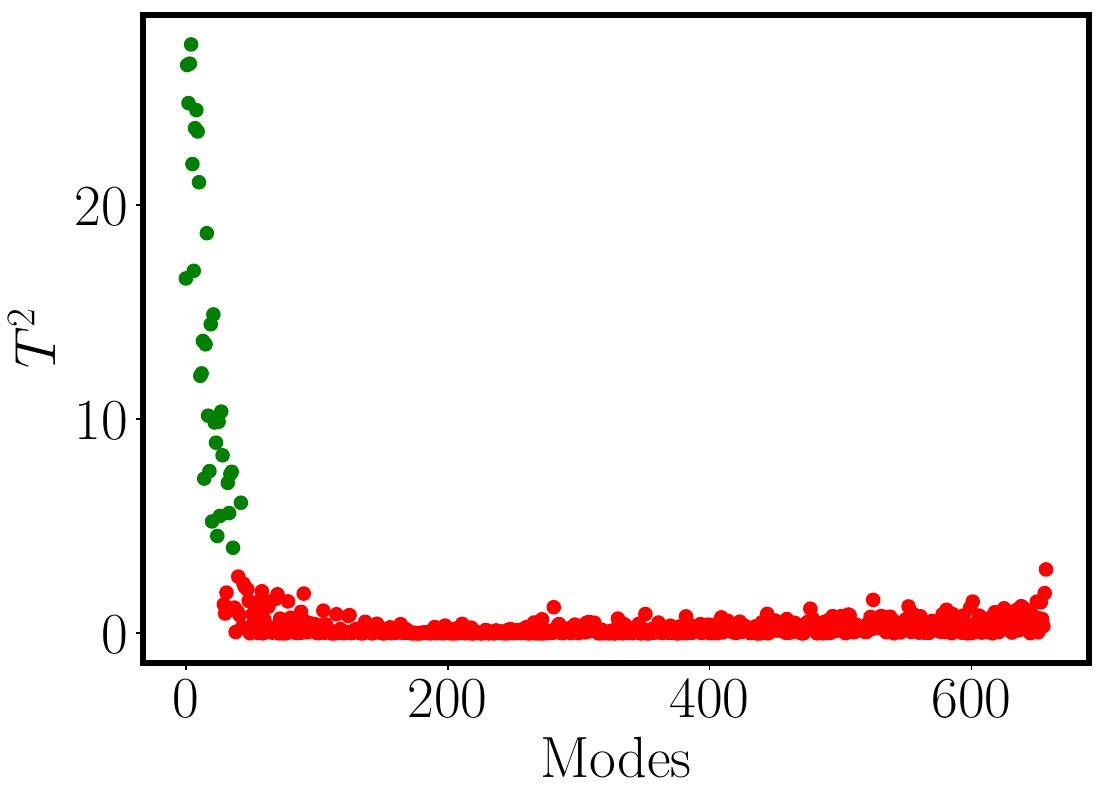}}
    \subfloat[]{\includegraphics[width=0.49\textwidth]{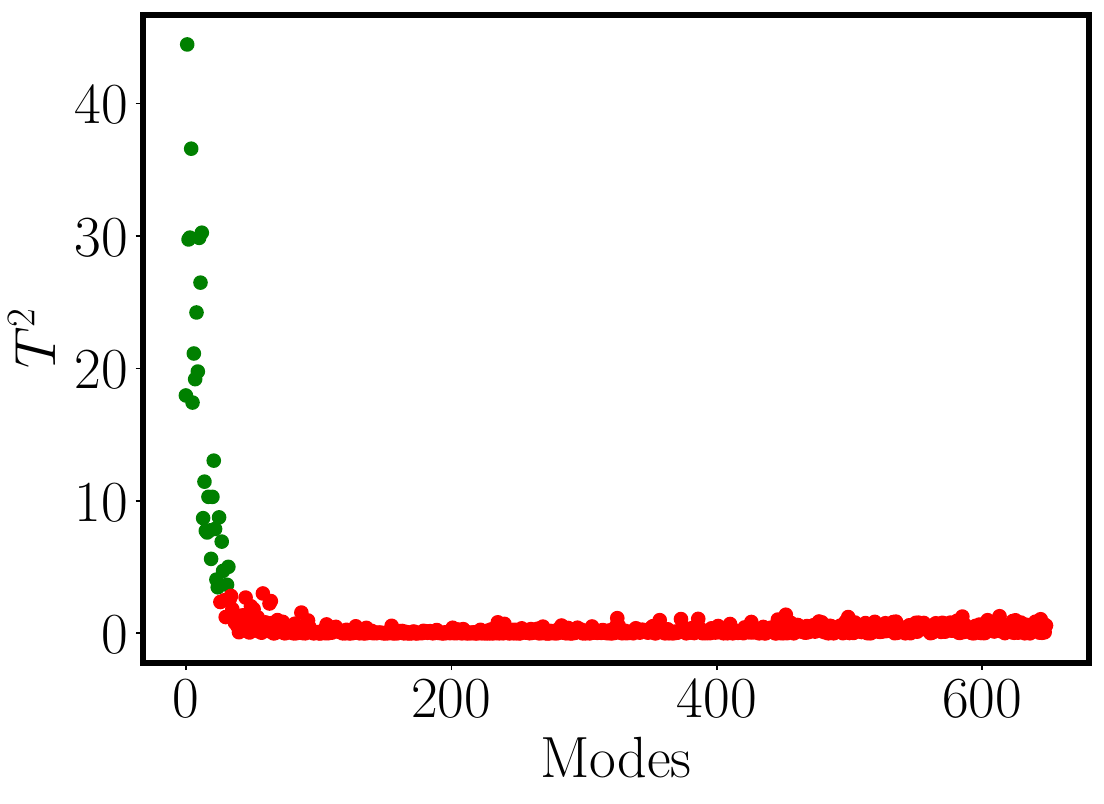}}
    \caption{$T^2$ clustering results for $\delta=2.5^{\circ}$ (a), $\delta=5^{\circ}$ (b), $\delta=10^{\circ}$ (c) and $\delta=12.5^{\circ}$ (d). Green dots represent the coherent modes and the red dots the non-coherent ones.\\}
    \label{fig:filtering}
    \vspace{0.5cm}
    \begin{tabularx}{0.75\textwidth}{Xcc}
        \toprule
        \textbf{$\delta$} & \textbf{Number of modes} & \textbf{Recovered energy} \\
        \midrule
        $2.5^{\circ}$ & 39 & 57.5\%  \\
        $5^{\circ}$ & 36 & 58.1\%  \\
        $10^{\circ}$ & 35 & 56.0\%  \\
        $12.5^{\circ}$ & 32 & 52.0\%  \\
        \bottomrule
    \end{tabularx}
    \captionof{table}{Number of coherent modes and total amount of energy recovered by them for each training angle.}
    \label{tab:modes}
\end{figure}

The first step to build the common basis is to perform the POD of each of the training angles individually. After that, the PSD-based mode-selection process described in \autoref{sec:methodology} is applied. \autoref{fig:filtering} shows the $T^2$ clustering results for each of the angles. The threshold for the coherent-mode selection is set to $T^2=3$. Table \ref{tab:modes} shows the number of selected modes in each case and the amount of energy recovered. It can be seen that the tendency is to have between 30 and 40 modes per angle containing coherent structures that represent a somewhat larger amount than half of the total energy. The case at $\delta=12.5^{\circ}$ is the one in which the coherent modes account for the smallest energy percentage, 52.0\%, while for $\delta=5^{\circ}$ they account for up to 58.1\%. It is important to note that the rest of the energy is shared among the remaining 600 non-coherent modes, therefore, each of them has a small individual contribution to the total energy of the system. 

\begin{figure}
    \centering
    \includegraphics[width=\textwidth]{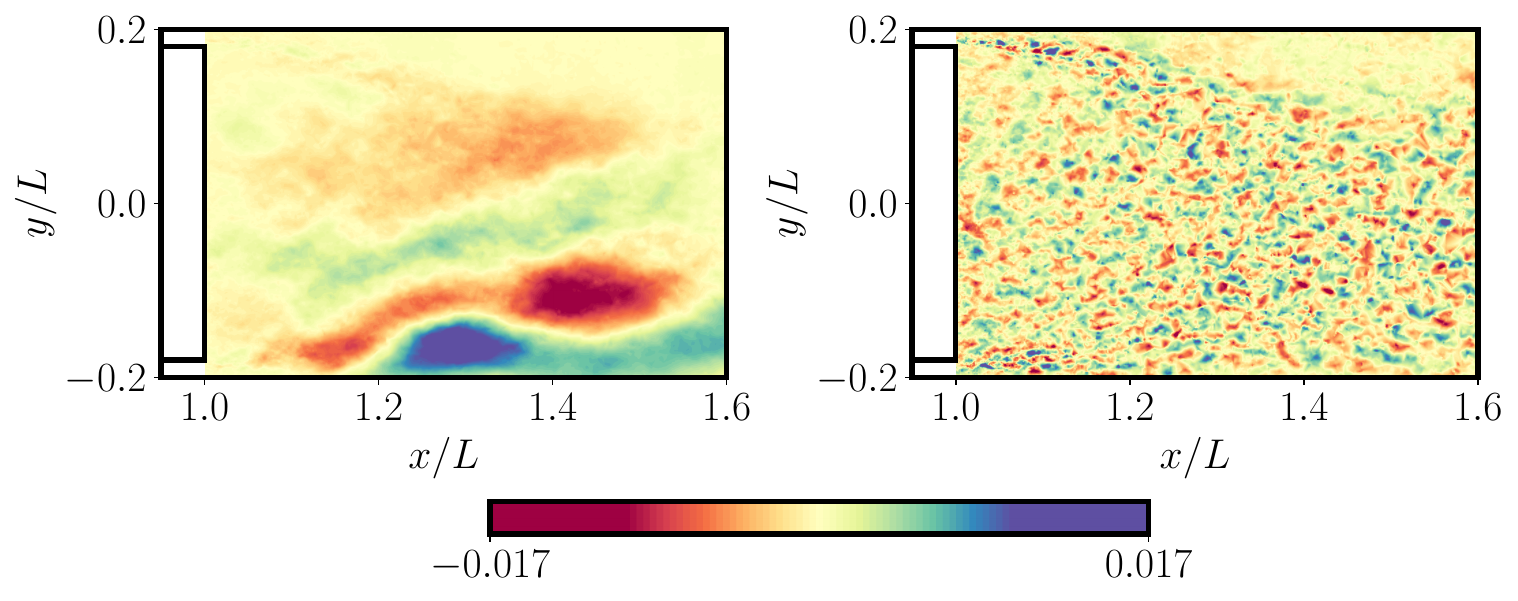}
    \caption{Spatial correlations of the 5th (left) and 450th (right) most energetic modes of $\delta=10^{\circ}$, clustered as coherent and non-coherent, respectively.}
    \label{fig:spatial_modes}
\end{figure}

Two modes of the case at $\delta=10^{\circ}$ are used to illustrate the clustering process. \autoref{fig:spatial_modes} compares the spatial correlation of a coherent mode with the one of a non-coherent mode. Note that the chosen coherent mode is the fifth most energetic one and the non-coherent mode is the 450th most energetic one. The coherent mode is clearly dominated by four large correlated regions linked to the vortex shed from the windward side of the vehicle, whereas the non-coherent mode depicts multiple small scales.

\begin{figure}
    \centering
    \subfloat[]{\includegraphics[width=\textwidth]{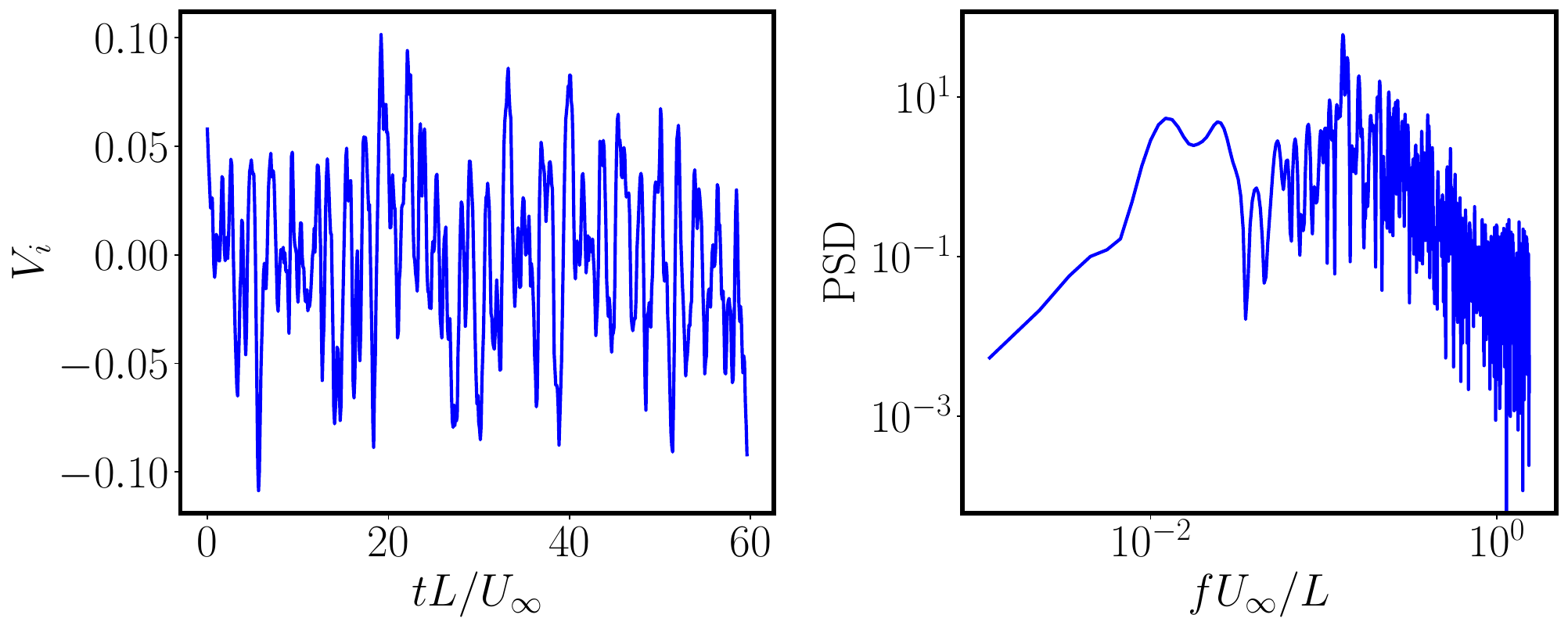} \label{fig:spec_coherent}}\\
    \subfloat[]{\includegraphics[width=\textwidth]{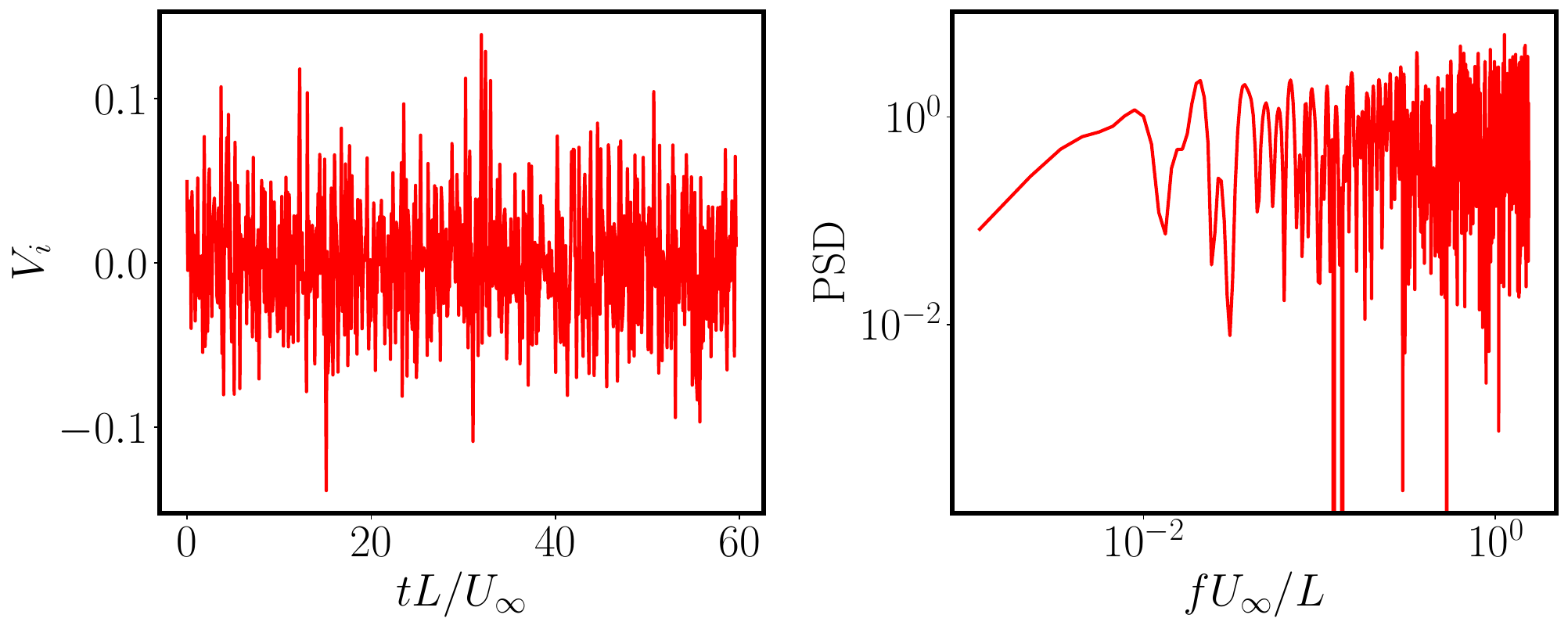}\label{fig:spec_non}}
    \caption{Temporal coefficient (left) and its power spectrum (right) of the 5th (a) and 450th (b) most energetic modes of $\delta=10^{\circ}$, clustered as coherent and non-coherent, respectively.}
    \label{fig:clustered}
\end{figure}

\autoref{fig:clustered} compares the temporal coefficient and its spectrum for both modes.  The spectrum of the coherent mode (\autoref{fig:spec_coherent} right) exhibits a peak at the non-dimensional frequency of $fH/U_{\infty}=0.13$. This peak corresponds to the windward vortex-shedding frequency \citep{eiximeno2024toward,BOOYSEN2022110562}. Note that no dominant frequencies can be observed in the spectrum of the non-coherent mode (\autoref{fig:spec_non} right), which is completely flat as those from pure white noise signals. These modes are seen as noise in the reduced system as their temporal coefficients are completely uncorrelated. The temporal coefficients of the non-coherent modes (\autoref{fig:spec_non} left) suggest that the lack of correlation might come from an inadequate sampling frequency, this one being lower than the dominant frequency of these modes. The noisy and random evolution of the non-coherent modes, together with their small individual energy contribution, would increase the cost of a surrogate model if they were included in the reduced system. However, they cannot be discarded without an efficient closure that accounts for the large energy percentage that they contain as a group.

\begin{figure}
    \centering
    \includegraphics[width=0.9\textwidth]{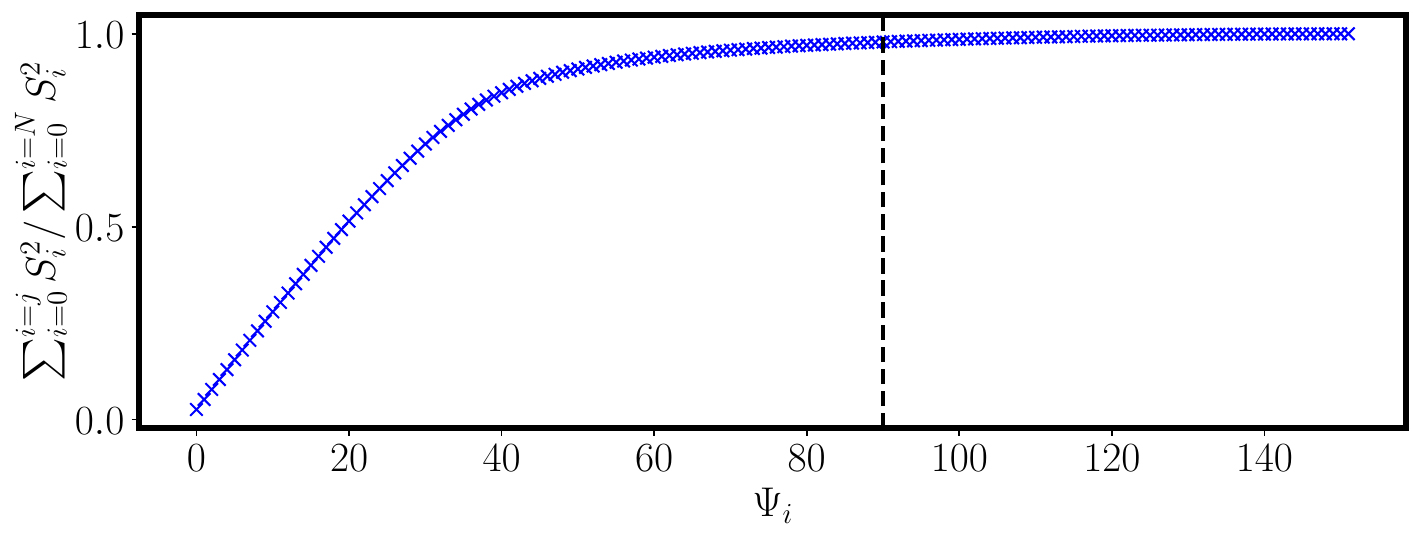}
    \caption{Cumulative energy for the POD to find the common basis between the selected modes.}
    \label{fig:cumcom}
\end{figure}

The spatial correlations of the selected modes are then concatenated to create the matrix $\mathbf{Y}$ as in \autoref{eqn:y}. As some coherent modes might be repeated in the yaw angle range, POD is applied to matrix $\mathbf{Y}$ to find the optimal and orthonormal basis that contains the information of the selected modes for the four angles. \autoref{fig:cumcom} shows the cumulative singular values to prove that instead of using all the 142 coherent modes, 90 vectors are enough to represent the information of all the coherent modes in the yaw-angle range under study. 

\begin{figure}
    \centering
    \subfloat[]{\includegraphics[width=0.5\textwidth]{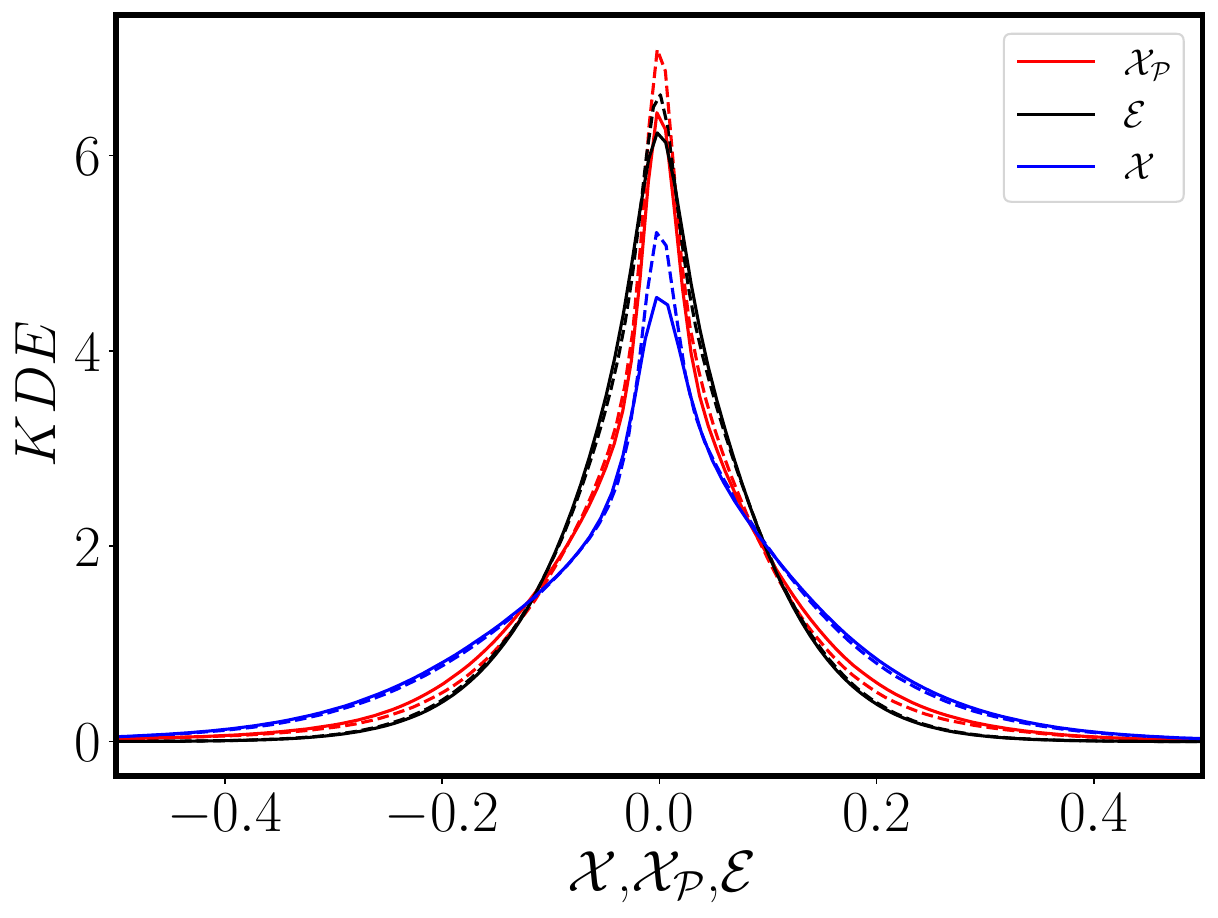}\label{fig:singlepdf}}
    \subfloat[]{\includegraphics[width=0.5\textwidth]{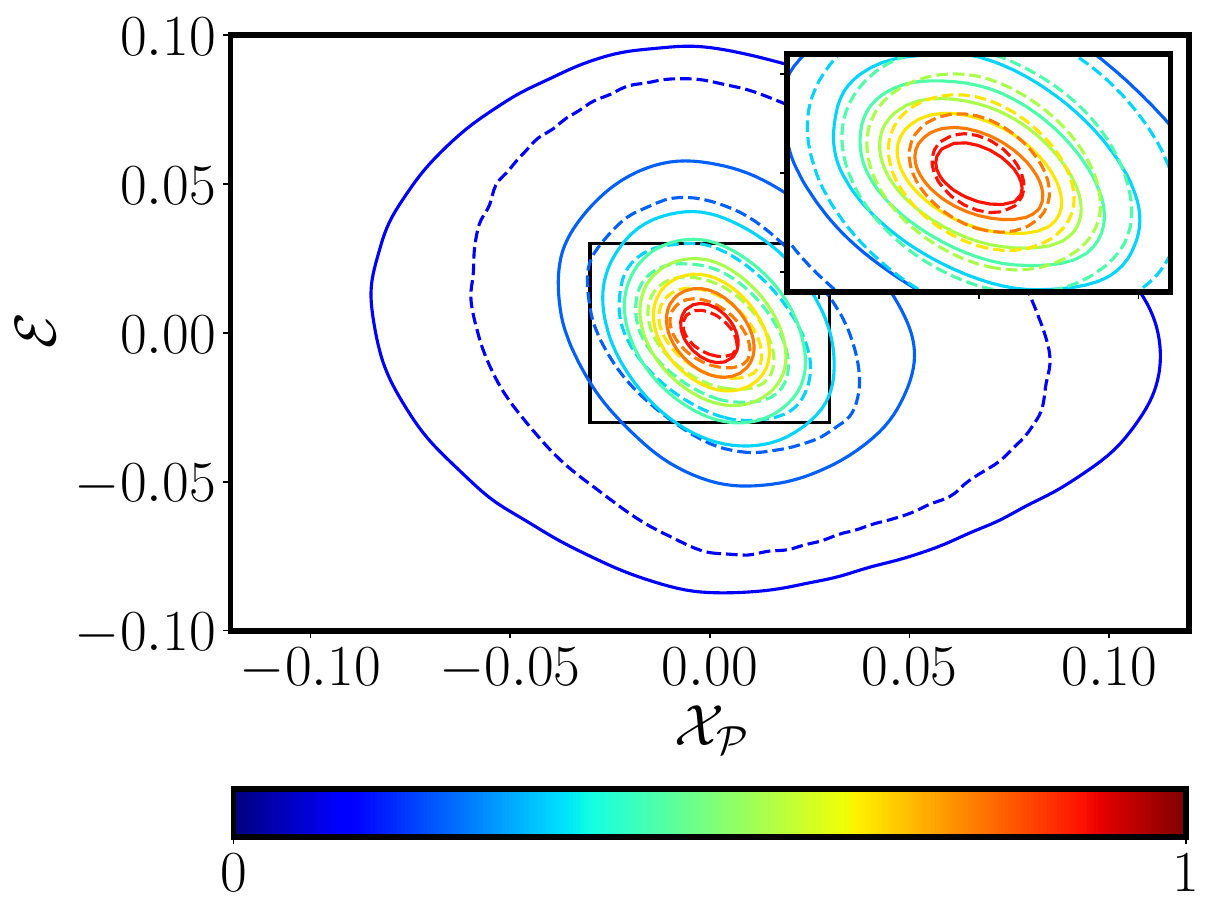}\label{fig:commonpdf}}
   \caption{Kernel density estimate of the velocity fluctuations, its reconstruction from the common basis and the error between both for all the snapshots in the training (solid) and test (dashed) datasets (a). Joint probability density function of the error depending on the reconstruction from the common basis, $p({\cal{E}}|{\cal{X_P}})$ for all the snapshots in the training (solid) and test (dashed) datasets.}
   \label{fig:common_single_pdf}
\end{figure}

\autoref{fig:singlepdf} presents the kernel density estimate of all the training snapshots for the original field, $\cal{X}$, its reconstruction after being projected into the common POD basis, $\cal{X_P}$, and the error between both of them, $\cal{E}$. The most likely situation is to have fluctuations close to zero in the original and reconstructed fields. This is explained by the large unperturbed area in the leeward ($y\geq 0$) side of the domain. The source of error is then the filtering performed by the POD reconstruction of the high-amplitude fluctuations. Such filtering yields a field that is more likely to have points with velocity fluctuations close to zero than in the original case. \autoref{fig:singlepdf} also confirms that this is holds true for the test case at $\delta=7.5^{\circ}$, bringing evidence that the common basis is valid for any angle in the studied range. \autoref{fig:commonpdf} compares the joint probability density function of the error given the reconstruction from the common basis, $p(\cal{E} \mid \cal{X}_{P})$, for the training and tests fields. As stated in \autoref{sec:methodology}, this is the probability density function learnt by the closure model as it ensures that the predicted error yields a statistically equivalent system to the original one. Consistently with the results shown in \autoref{fig:singlepdf}, the most likely case in both the training and test datasets is to have a state with the velocity reconstruction and its error with the original field being close to zero. The most probable values for the test set match the ones of the training set, however, the limits of $p(\cal{E} \mid \cal{X}_{P})$ for the training set are wider than those at $\delta=7.5^{\circ}$. 

\subsection{Statistical closure accuracy}
\begin{figure}
\centering
\includegraphics[width=\textwidth]{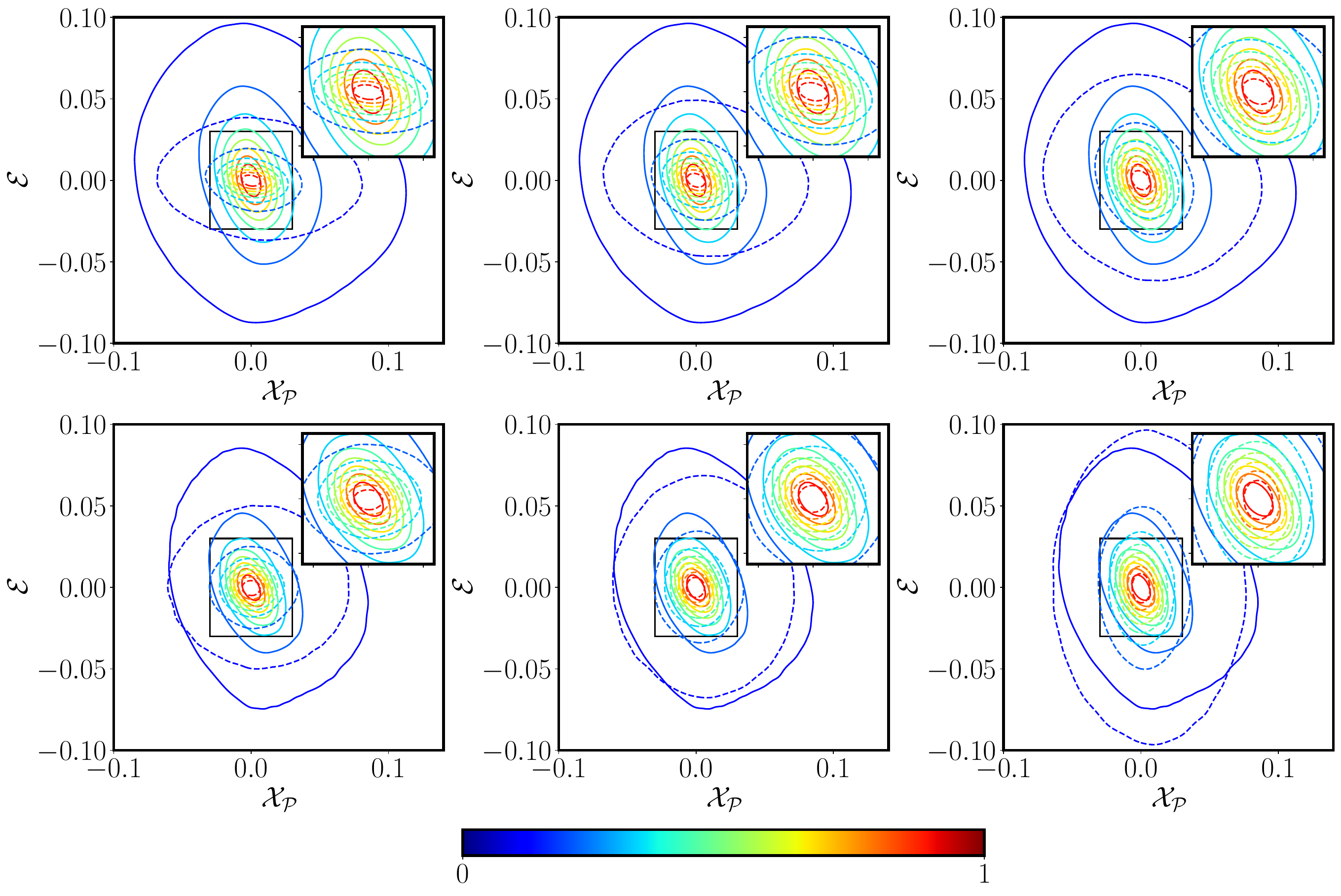}
\caption{Joint probability density function of the error depending on the reconstruction from the common basis, $p({\cal{E}}|{\cal{X_P}})$. Solid lines represent the reference values and dashed lines represent the values learnt by the closure. The figures above represent the accuracy on the training set ($\delta=[2.5^{\circ}, 5^{\circ}, 10^{\circ}, 12.5^{\circ}]$)  and the figures below show the accuracy for the validation set ($\delta=7.5^{\circ}$). Architecture 1 is on the left, architecture 2 at the center and architecture 3 on the right.\\}
\label{fig:training_pdfs}
\vspace{0.5cm}
\begin{tabularx}{0.6\textwidth}{Xcc}
    \toprule
    \textbf{Attention size ($d_{\text{model}}$)} & \textbf{${\cal{{D_{KL}}}}$ Training} & \textbf{${\cal{{D_{KL}}}}$ Test} \\
    \midrule
    64  & 0.0159 & 0.0057  \\
    128 & 0.0106 & 0.0016  \\
    256 & 0.0056 & -0.0017  \\
    \bottomrule
\end{tabularx}
\captionof{table}{Küllback-Leibler divergence between the original $p({\cal{E}}|{\cal{X_P}})$ and the one learnt by each transformer architecture.}
\label{tab:kltraining}
\end{figure}

The three different architectures described in \autoref{tab:architectures} are tested in order to assess the correct size of the attention layer. In \autoref{fig:training_pdfs}, the probability density function, $p({\cal{E}}|{\cal{X_P}})$, given by the transformer output with the original one, represented in \autoref{fig:commonpdf}, for both the training and test datasets are compared. Architecture 1, with an attention layer of size $d_{\text{model}}=64$, performs poorly in learning both the center and the limits of the distribution. The Kullback--Leibler divergence between the transformer prediction and the original probability for the training data is of ${\cal{{D_{KL}}}}=0.0159$ and for the test data is ${\cal{{D_{KL}}}}=0.0057$. Both values are the highest ones obtained during the architecture refinement process. In this case, the main source of error is that the PDF learnt by the transformer is much narrower than the original one, meaning that the model fails to recover the fluctuations with larger amplitude.

Increasing the attention layer to $d_{\text{model}}=128$, with its subsequent duplication of the number of parameters, allows the transformer to learn a wider area of $p({\cal{E}}|{\cal{X_P}})$. This reduces the KL divergence with the original data to ${\cal{{D_{KL}}}}=0.0106$ for the training snapshots and ${\cal{{D_{KL}}}}=0.0016$ for the test snapshots. It is relevant to mention that this architecture nearly matches the output distribution for the test set as the limits of $p({\cal{E}}|{\cal{X_P}})$ for $\delta=7.5^{\circ}$ are narrower than the ones found in the training set.

Duplicating the attention size to $d_{\text{model}}=256$ leads to the best match of the training dataset of the three architectures. The learnt PDF expands for a wider area of fluctuations and the KL divergence is reduced to ${\cal{{D_{KL}}}}=0.0056$. Now the KL divergence on the test set is ${\cal{{D_{KL}}}}=-0.0017$. The negative sign accounts for the larger fluctuations from the training set that are not present in the case of $\delta=7.5^{\circ}$ and are already learnt by the transformer. Moreover, in this case the absolute value of ${\cal{{D_{KL}}}}$ is slightly larger than the one found with architecture 2. This is the last step of architecture refinement because the evaluation of the test set has already crossed the ideal prediction in which the KL divergence would be null as the model shows the firsts signs of overfitting.

\begin{figure}
    \centering
    \includegraphics[width=\textwidth]{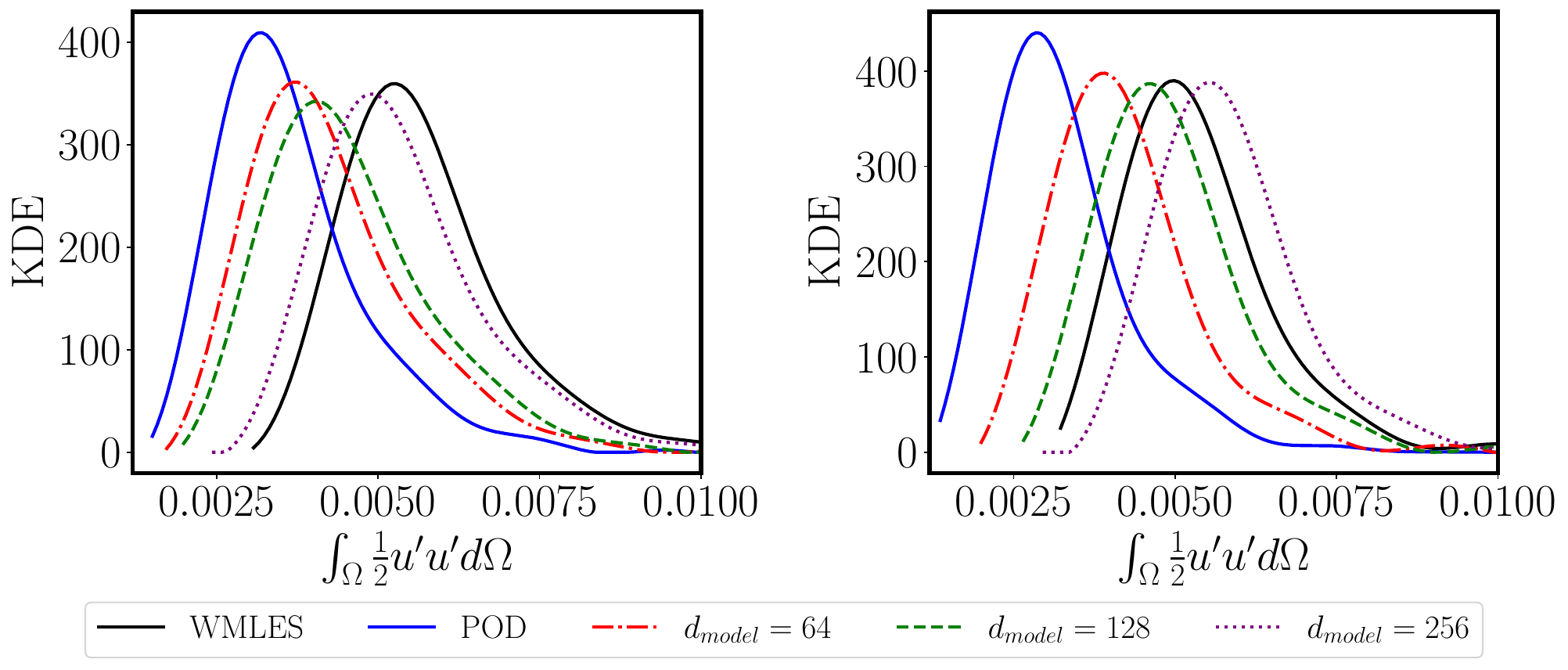}
    \caption{Kernel density estimate of the energy for the training (left) and tests (right) datasets}
    \label{fig:kde_energy}
\end{figure}

The wider area of $p({\cal{E}}|{\cal{X_P}})$ learnt by the architectures with larger attention size can be linked to the amount of turbulent kinetic energy (TKE), 
\begin{equation}
    k = \int_{\Omega}\frac{1}{2}u'u' d\Omega,
\end{equation}
recovered by the closure model. The TKE recovered in each case is quantified with the kernel density estimate among all the snapshots of the training and test sets separately. \autoref{fig:kde_energy} effectively showcases that the most likely energy value, $\overline{k}$, after the reconstruction from the POD common basis is significantly lower than the one of the original flow. For the training snapshots, it is reduced from $\overline{k}=0.0053$ to $\overline{k}=0.0032$ and for the test dataset it decreases from $\overline{k}=0.0050$ to $\overline{k}=0.0029$. 

\autoref{fig:kde_energy} also shows that the most likely energy value when adding the closure term increases with the attention layer size of the transformer used to model the missing fluctuations. In the training angles, $\overline{k}$ increases from $\overline{k}=0.0038$ to $\overline{k}=0.0041$ when the attention sizes changes from $d_{\text{model}}=64$ to $d_{\text{model}}=128$. It finally reaches the value of $\overline{k}=0.0049$ with the largest architecture of $d_{\text{model}}=256$. A similar behavior is observed with the case at $\delta=7.5^{\circ}$, for the architectures with $d_{\text{model}}=64$ and $d_{\text{model}}=128$ as $\overline{k}$ goes up to $\overline{k}=0.0039$ and $\overline{k}=0.0046$, respectively. However, for the architecture with $d_{\text{model}}=256$ the most likely energy value, $\overline{k}=0.0053$, is slightly higher than in the original flow. This can be explained by the fact that the probability density function of the fluctuations predicted by the transformer is wider than the ones of the real case (\autoref{fig:training_pdfs}).

This analysis brings evidence that the closure model actually reduces the offset between the energy of the POD reconstruction and the one of the original system.
It is important to note that the accuracy on the energy prediction is directly linked with the KL divergence between the predicted $p({\cal{E}}|{\cal{X_P}})$ by the transformer and the ground truth. When the KL divergence is positive, the energy added by the closure is still smaller than the gap between the original flow and the POD reconstruction. Zero KL divergence would mean a perfect match between the model and the ground truth with no energy deviation. On the last scenario, a negative KL divergence indicates that the model is overshooting the predicted fluctuations, and with it the turbulent kinetic energy. Either with a positive or negative KL divergence, a larger absolute value indicates a larger deviation in the additional turbulent kinetic energy.

\subsection{Spatial field reconstruction}

Up to this point of the discussion, it has been proven that adding a field of fluctuations based on the $p({\cal{E}}|{\cal{X_P}})$ learnt by the proposed transformer architectures is enough to close the energy gap of a POD reconstruction and the original flow field. However, it remains to be proven that the closure model can distribute these fluctuations adequately across the spatial and temporal domains.

\begin{figure}
    \centering
    \subfloat[]{\includegraphics[width=\textwidth]{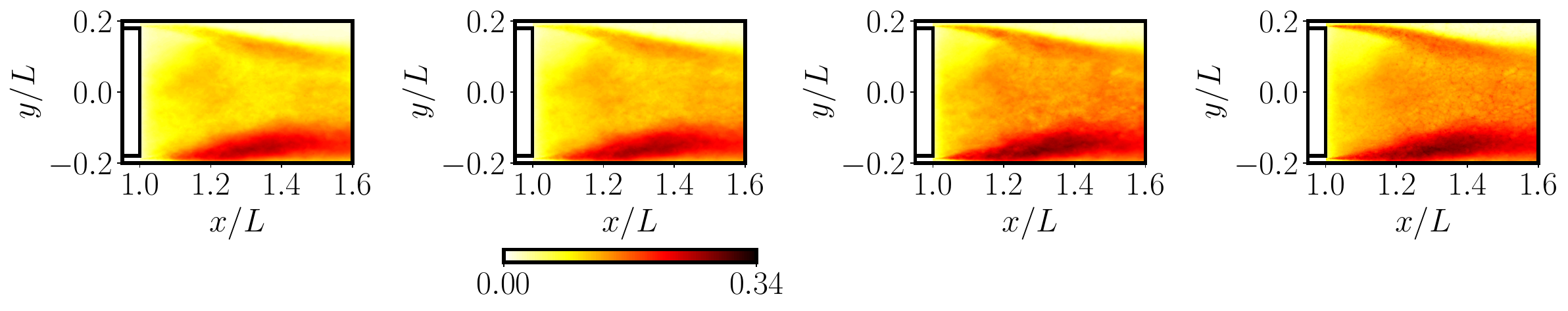}\label{fig:contour_train}}\\
    \subfloat[]{\includegraphics[width=\textwidth]{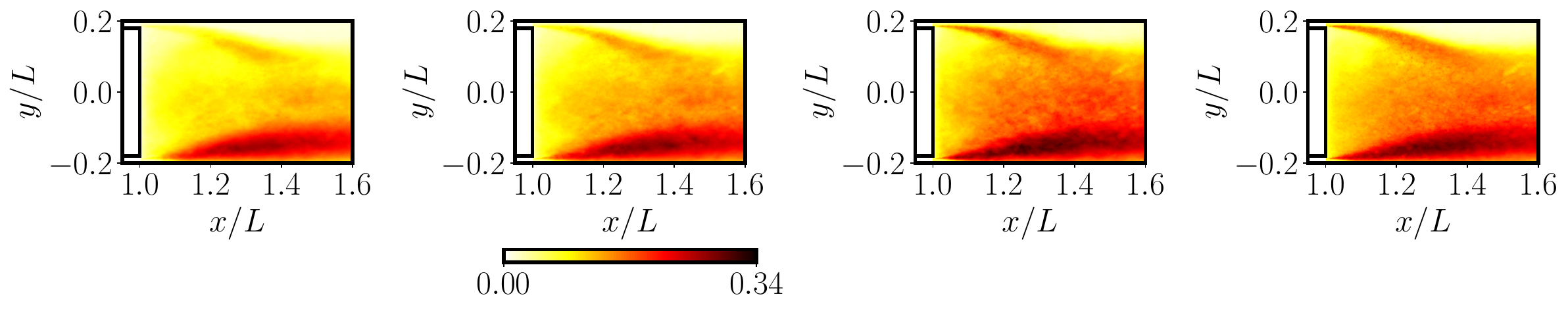}\label{fig:contour_test}}
   \caption{Root mean square value of the streamwise velocity fluctuations for the cases at $\delta=2.5^{\circ}$ (a) and $\delta=7.5^{\circ}$. From left to right, the pictures represent: reconstruction from the common POD basis, reconstruction with the closure model with an attention size of $d_{\text{model}}=64$, reconstruction with the closure model with an attention size of $d_{\text{model}}=256$ and the original flow}
   \label{fig:contour_architectures}
\end{figure}

\begin{figure}
    \centering
    \subfloat[]{\includegraphics[width=0.5\textwidth]{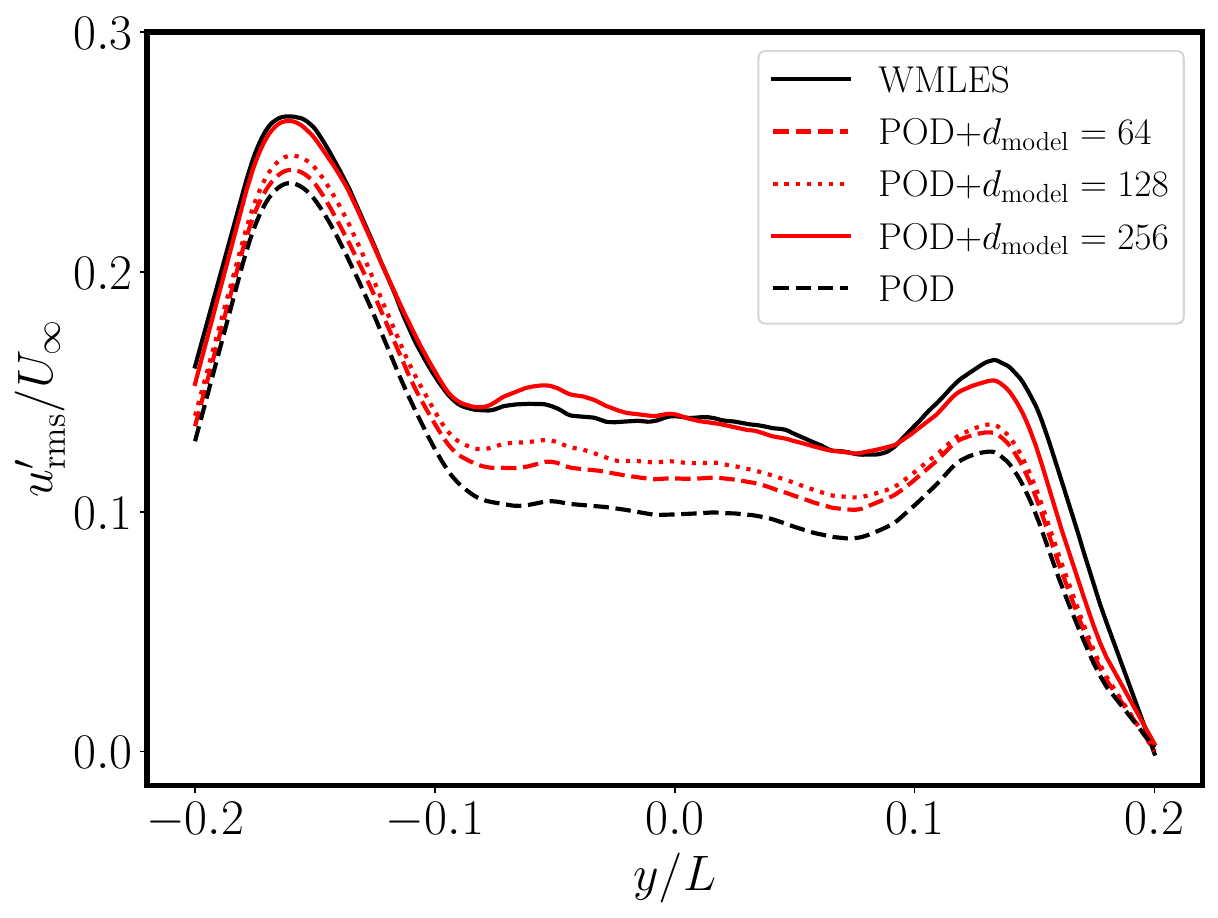}\label{fig:line_train}}
    \subfloat[]{\includegraphics[width=0.5\textwidth]{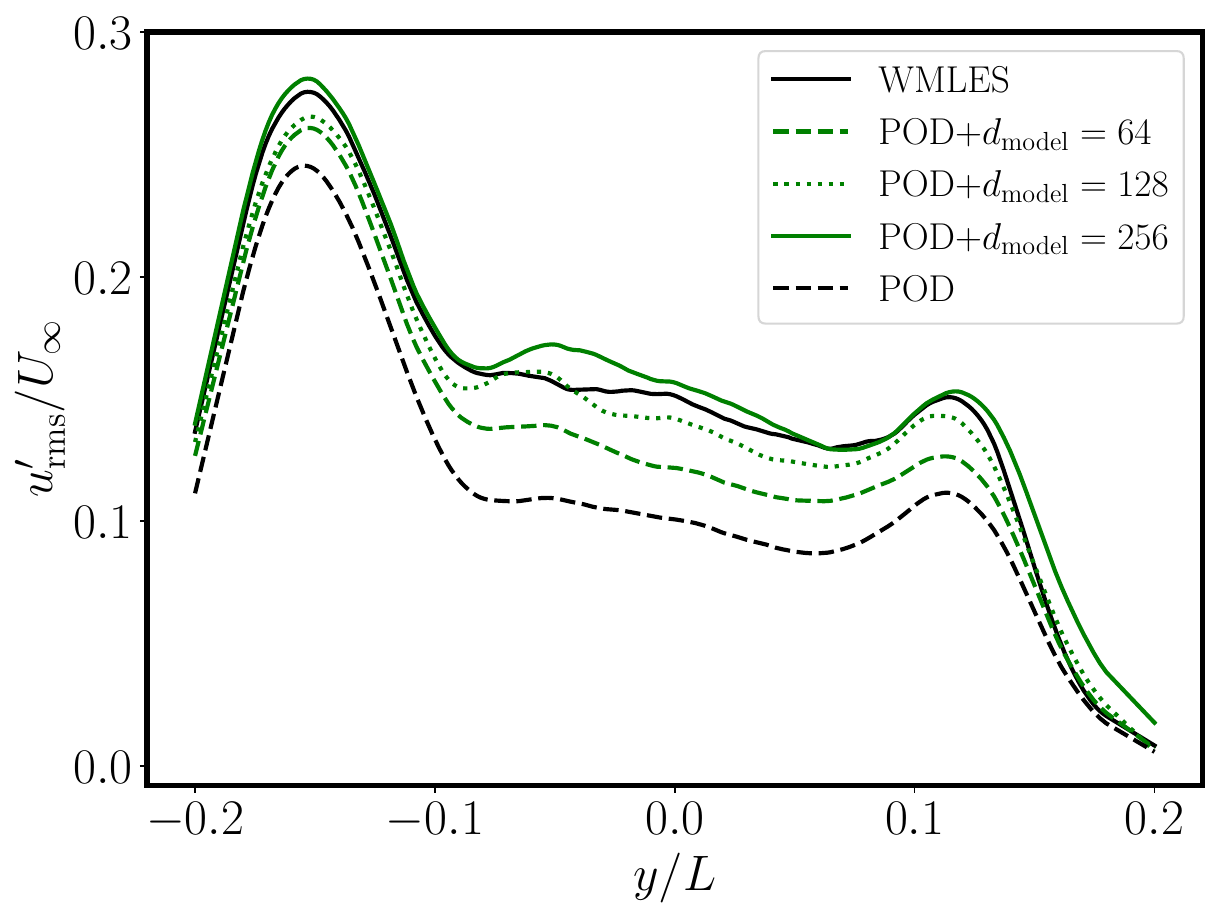}\label{fig:line_test}}
   \caption{Root mean square of the streamwise velocity fluctuations for the cases at $\delta=2.5^{\circ}$ (a) and $\delta=7.5^{\circ}$ (b) on a cross-stream line at $x/L=1.3$.}
   \label{fig:line_architectures}
\end{figure}

\autoref{fig:contour_architectures} shows the root mean square (rms) of the velocity fluctuations in the spatial domain for the reconstruction from the common POD basis, the closure term trained with $d_{\text{model}}=64$ and $d_{\text{model}}=256$ together with the ones of the original field. The rms of velocity fluctuations is closely related with the local contribution to the total turbulent kinetic energy of the flow, hence, a field with the closure term matching the rms fluctuations of the original case could be considered accurate in space and statistically equivalent in time. In the figure, the case at $\delta=2.5^{\circ}$ is used to illustrate the performance on the various training angles. The results at $\delta=7.5^{\circ}$ are also plotted to show the performance of the model on unseen data. Moreover, the four mentioned reconstructions together with the one at $d_{\text{model}}=128$ are evaluated along the line at $x/L=1.3$ in \autoref{fig:line_architectures}. The reader is referred to \autoref{appA} for the equivalent figures to \autoref{fig:contour_architectures} and \autoref{fig:line_architectures} corresponding to the training cases at $\delta=[5^{\circ}, 10^{\circ}, 12.5^{\circ}]$.

Both figures illustrate that the common basis captures the positions of the fluctuation maxima and their correct distribution along the domain, ensuring that the main flow structures are preserved throughout the projection and reconstruction processes (\autoref{eqn:project} and \autoref{eqn:reproject}). This is also valid for the case at $\delta=7.5^{\circ}$, despite its features were not explicitly included in the basis.

In all analyzed angles, the reconstruction from the common basis misses the actual value by an offset associated with the filtered fluctuations. \autoref{fig:contour_architectures} and \autoref{fig:line_architectures} show that increasing the attention size helps to close the gap in the rms fluctuations in all areas of the domain. The larger range of fluctuations learnt by the deeper architecture ($d_{\text{model}}=256$) and its additional kinetic energy added to the flow, translates to a nearly perfect match of the rms of the velocity fluctuations. It is worth mentioning that in the training cases most of the differences between the original field and the closure prediction arise from the model underestimating the fluctuations, however, at $\delta=7.5^{\circ}$ all the error of the closure is attributed to a slight overprediction. 

As the offset between the POD reconstruction and the original field is not constant throughout the whole domain, \autoref{fig:contour_architectures} and \autoref{fig:line_architectures} are also the evidence that the closure learns how much energy the POD reconstruction missed depending on the domain region. This proves that the closure model does not only close the energy gap statistically over all the points of all snapshots, but also that it can give accurate predictions of what happens in every point in the domain.

\subsection{Instantaneous-field reconstruction}
After discussing how the closure model can emulate a field which is statistically equivalent to the original one in all points of the domain, it is time to discuss its impact on the reconstruction of the instantaneous fields. In this case, all comparisons are done with architecture 3 ($d_{\text{model}}=256$) as it is the only one able to close all the energy gap between the reconstruction and the original flow.

\begin{figure}
    \centering
    \includegraphics[width=0.7\textwidth]{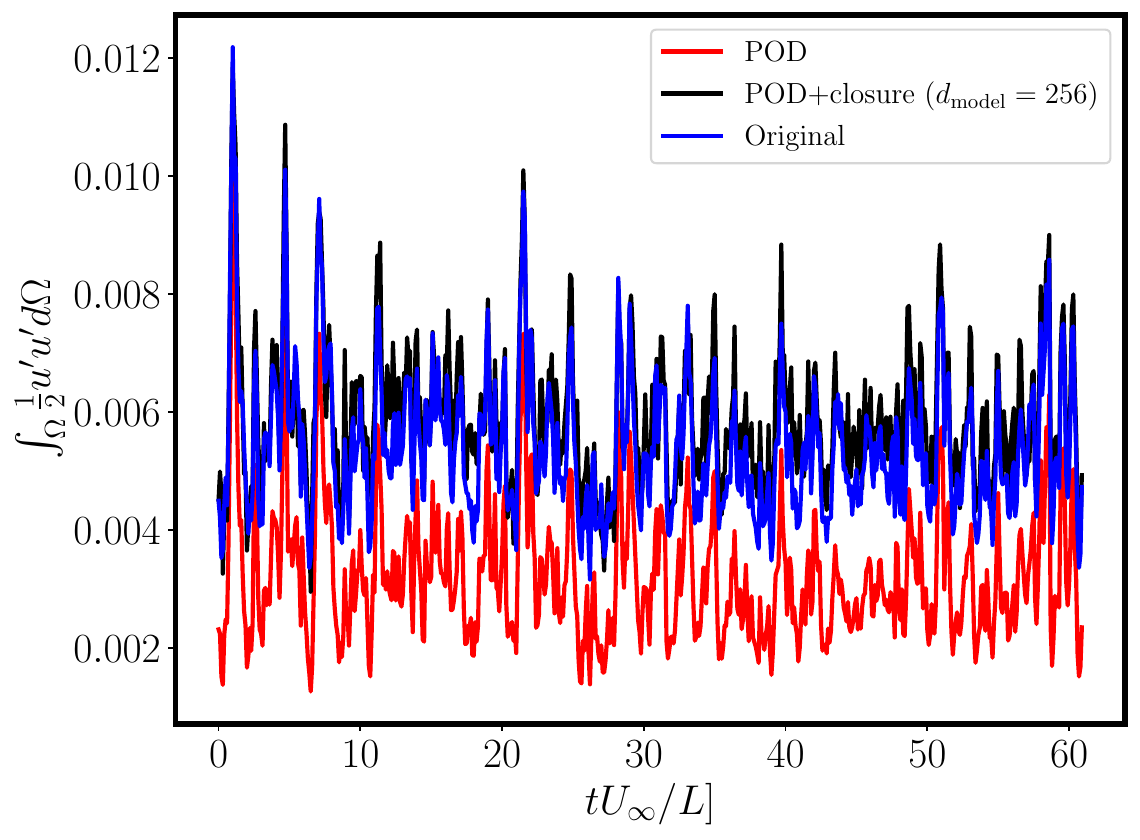}
    \caption{Temporal evolution of the energy of the system for the case at $\delta=7.5^{\circ}$\\}
    \label{fig:tempenergy}
    \vspace{0.5cm}
    \begin{tabularx}{0.7\textwidth}{Xcc}
        \toprule
        \textbf{$\delta$} & \textbf{POD reconstruction} & \textbf{Closure ($d_{\text{model}}=256$)} \\
        \midrule
        $2.5^{\circ}$ & 38.6\% & 8.2\%  \\
        $5^{\circ}$ & 37.0\% & 8.8\%  \\
        $7.5^{\circ}$ & 41.8\% & 10.7\%  \\
        $10^{\circ}$ & 38.5\% & 7.1\%  \\
        $12.5^{\circ}$ & 40.4\% & 11.4\%  \\
        \bottomrule
    \end{tabularx}
    \captionof{table}{Mean relative error between the energy of the original field, the energy recovered by the POD reconstruction and the POD reconstruction with the closure model.}
    \label{tab:energy}
\end{figure}

The first step is proving that the closure learns the amount of energy missing in each timestep. To do so, \autoref{fig:tempenergy} shows the temporal evolution of the total turbulent kinetic energy together with its POD reconstruction and the reconstruction corrected with the closure model for the case at $\delta = 7.5^{\circ}$. This case is the evidence that common POD basis successfully captures the instants of all energy maxima and minima of the original field. Once again, this is still valid even if the flow condition was not included in the database.

\autoref{fig:tempenergy} also shows that the closure model represents the energy missing in each snapshot instead of adding the same energy to all of them. However, in the particular case of $\delta = 7.5^{\circ}$, the actual energy predicted by the closure is consistently higher than the one of the original flow in each snapshot, as have been discussed in the previous paragraphs. Table \ref{tab:energy} links the better prediction of the energy temporal evolution with the mean relative error regarding the energy of the original field. In all angles this error has been reduced from over $37\%$ to a margin between $7\%$ and $12\%$.

\begin{figure}
    \centering
    \includegraphics[width=\textwidth]{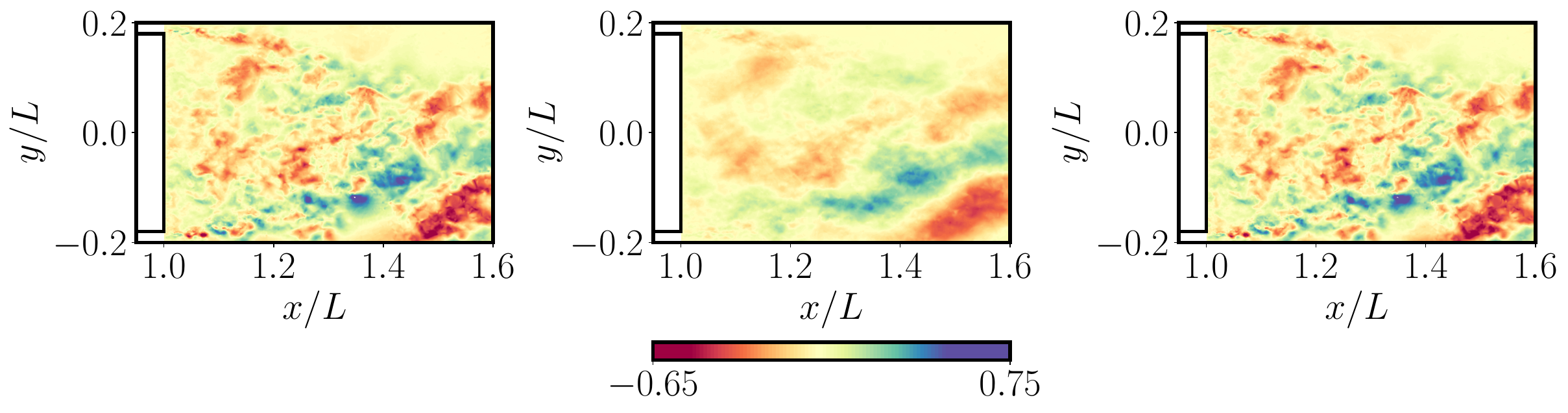}
    \caption{Original streamwise velocity fluctuations (left), POD reconstruction (center) and POD reconstruction with the closure term (right) for a snapshot at $\delta=7.5^{\circ}$.}
    \label{fig:instant_comparison}
\end{figure}

\begin{figure}
    \centering
    \subfloat[]{\includegraphics[width=0.5\textwidth]{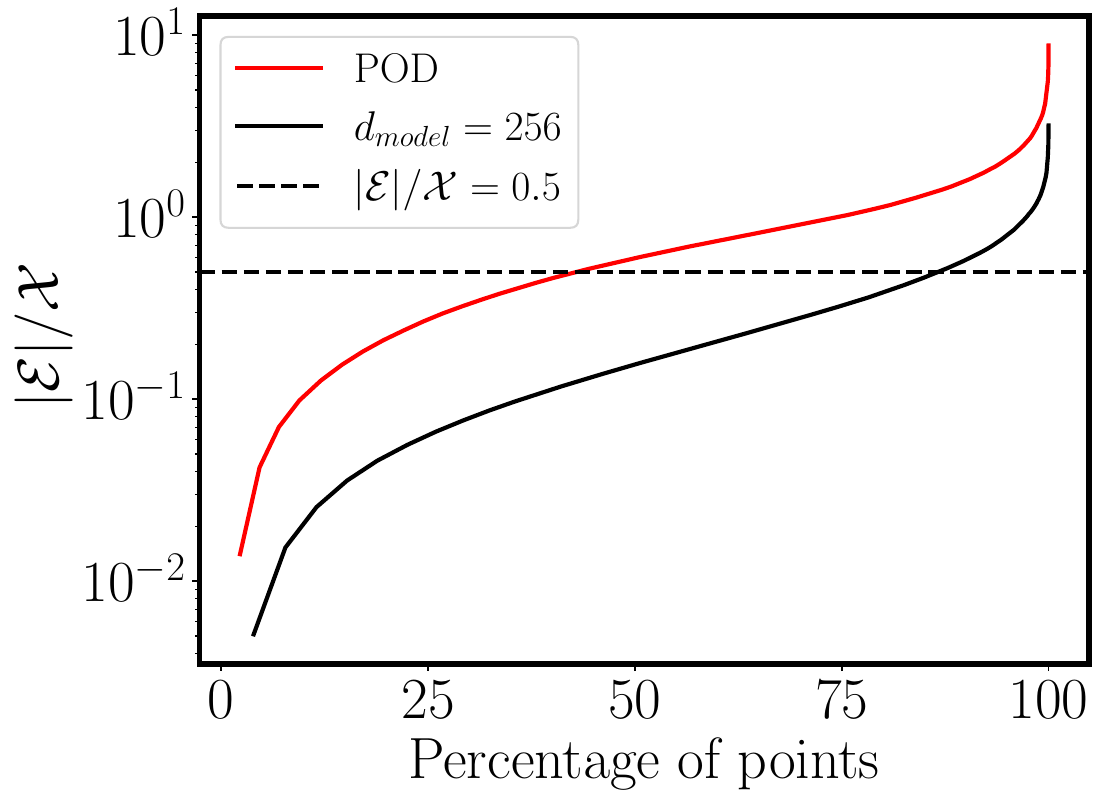}\label{fig:percentage_closed}}
    \subfloat[]{\includegraphics[width=0.5\textwidth]{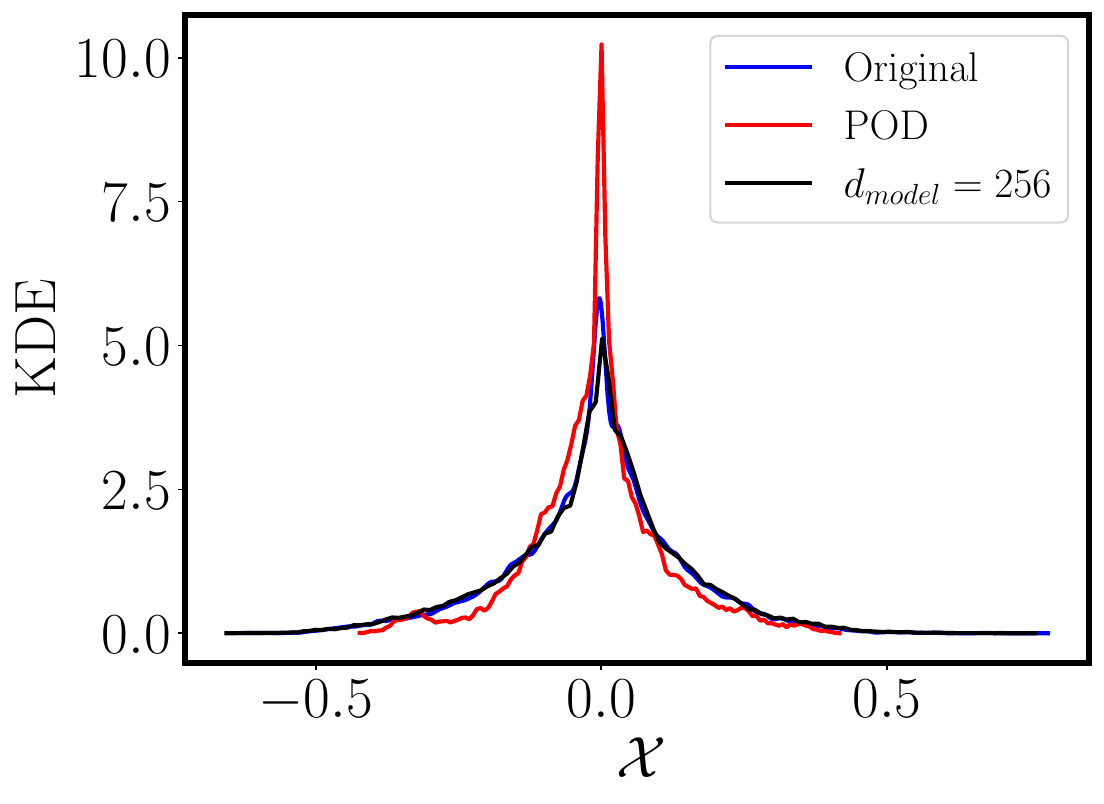}\label{fig:pdf_closed}}
    \caption{Cummulative density function of the relative error between the reconstructions and the original field (a) and probability density function for the three fields (b).}
    \label{fig:closed_all}
    \vspace{0.5cm}
    \begin{tabularx}{0.6\textwidth}{Xcc}
        \toprule
        \textbf{$\delta$} & \textbf{POD reconstruction} & \textbf{Closure ($d_{\text{model}}=256$)} \\
        \midrule
        $2.5^{\circ}$ & 0.2166 & 0.0240  \\
        $5^{\circ}$ & 0.2085 & 0.0269  \\
        $7.5^{\circ}$ & 0.2362 & 0.0253  \\
        $10^{\circ}$ & 0.2111 & 0.0222  \\
        $12.5^{\circ}$ & 0.2062 & 0.0245  \\
        \bottomrule
    \end{tabularx}
    \captionof{table}{Kullback--Leibler divergence, $\cal{D_{KL}}$, between the original field and the reconstruction with and without the closure term. The results are averaged over all snapshots for each angle.}
    \label{tab:invent}
\end{figure}

Closing the energy gap appropiately in each snapshot also comes with a better prediction of the instantaneous fluctuations. To exemplify this, \autoref{fig:instant_comparison} compares the original instantaneous field and its reconstruction for a handpicked snapshot at $\delta=7.5^{\circ}$, effectively showcasing that the POD reconstruction exhibits large deviations from the original data. In fact, \autoref{fig:percentage_closed} illustrates that the reconstruction from the standard POD basis leads to a relative error larger than $| {\cal{E}}|/{\cal{X}} \geq 0.5$ in 57.8\% of the points in the domain. After adding the closure term, the accuracy of the reconstruction is increased so that only 13.7\% of the points have a relative error higher than $|{\cal{E}_M|/X}\geq 0.5$.

The comparison between the probability density functions (PDF) of the fields, \autoref{fig:pdf_closed}, agrees with \autoref{fig:singlepdf} on showing that the POD reconstruction filters the high-amplitude fluctuations by increasing the points with fluctuations close to zero. When adding the closure term, the probability density function of the velocity fluctuations is nearly identical to the one of the original field. In fact, the Kullback--Leibler (KL) divergence between the reconstructed PDFs and the original one is reduced from ${\cal{D_{KL}}} = 0.2447$ to  ${\cal{D_{KL}}} = 0.0052$ after adding the term predicted by the transformer. Table \ref{tab:invent} shows the mean ${\cal{D_{KL}}}$ over all snapshots to show that adding the closure term reduces the KL divergence with the original field of instantaneous velocity fluctuations regardless of the yaw angle.

\section{Conclusions}
\label{sec:conclusions}

This manuscript presents a deep-learning-based closure model for truncated POD modes. The main objective is to provide a methodology to recover the energy lost when building a surrogate model on a low dimensional space. To do so, a transformer model is used to learn the spatial probability density function of the difference between the original flow field and the POD reconstruction from the modes that would be included in a surrogate. The methodology is tested for the streamwise velocity fluctuations on a slice in the wake of the Windsor body at the yaw angles of $\delta=[2.5^{\circ}, 5^{\circ}, 7.5^{\circ}, 10^{\circ}, 12.5^{\circ}]$. As the model has to be generelizable for unseen data, the case at $\delta=7.5^{\circ}$ is used as a test dataset and the rest of angles are used for training.

Before working on the transformer model, a set of POD modes at each of the training angles is selected. Those modes have to be the most meaningful ones in the system as they would constitute the core of a reduced-order model. The selection process is based on performing principal-component analysis on the power-spectral density of the temporal coefficients. Then the modes with an outstanding frequency behavior are clustered with Hotelling's $T^2$. Those modes are named as coherent modes and this selection process ensures that they are the only ones that present relevant frequency dynamics.

In the particular case of the Windsor body, less than ten percent of the modes are coherent, however, they account for nearly 60\% of the energy. The remaining 40\% is distributed along the more than 600 non-coherent modes and is the one that needs to be modelled by the closure. The clustered modes from each training angle are concatenated to form a common basis that preserves the coherent structures inside the studied yaw angle range. POD is applied to the concatenated modes to ensure that all vectors from the basis are orthonormal between themselves and that they are the optimal representation of the coherent modes in that range. This operation reduced the number of basis vectors from 142 to 90 without any additional information loss.

Projecting any snapshot (regardless of whether it was included in the training set or not) into the common POD basis, filters the high-amplitude fluctuations. A transformer-encoder block with an easy-attention mechanism is used to learn the probability density function of the missing fluctuations depending on the reconstructed value from the POD common basis. Three different transformer architectures are trained in order to assess its effect on the recovered fluctuations. The main difference between the architectures is the change on the attention size. In the shallowest architecture it takes the value of $d_{\text{model}}=64$. Then it is doubled twice to get an attention size of $d_{\text{model}}=128$ and $d_{\text{model}}=256$.

The larger the attention size, the more fluctuations from the training set are recovered. The accuracy of the prediction is quantified with the KL divergence between the transformer output and the original field. For the training set it reduces from ${\cal{{D_{KL}}}}=0.0159$ to ${\cal{{D_{KL}}}}=0.0056$ when the attention size changes from $d_{\text{model}}=64$ to $d_{\text{model}}=256$. In the case of the test set, there is also an accuracy improvement when doubling the attention size up to $d_{\text{model}}=128$, but then, the first signs of overfitting to the training data are seen with the deepest architecture. The evaluation of the closure at $\delta=7.5^{\circ}$ for the architecture with $d_{\text{model}}=256$ is the only case in which the KL divergence has a negative value, ${\cal{{D_{KL}}}}=-0.0017$. Note that a negative KL implies that the transformer has learnt larger fluctuation amplitudes from the training set that are not present in the test set.

Adding the fluctuations field predicted by the transformer reduces the energy gap between the POD reconstruction and the original field. The architectures with larger attention size recover more energy than the shallower transformers. For instance, in the training set, the most likely energy value after adding the closure with $d_{\text{model}}=64$ is of $\overline{k}=0.0038$, and it rises to $\overline{k}=0.0049$ for $d_{\text{model}}=256$. In this case, since the fluctuations from the training set are larger than the ones of the test set, the overfitting observed when comparing the probability density function leads to an overshoot in the energy prediction. The evaluation of the test set with $d_{\text{model}}=256$ yields $\overline{k}=0.0053$ but the most likely energy value in the original flow is $\overline{k}=0.0050$.

Adding these fluctuations also leads to an improvement on the prediction of the root mean square value of the velocity fluctuations. The reconstruction from the common POD basis is able to capture the distribution of all local maxima and minima, but it falls short when matching the correct value. Then, the closure model helps to recover the missing fluctuations in the correct part of the domain. Once again, an increase on the attention size leads to a better closure of the offset. The evaluation of the deepest architecture at $\delta=7.5^{\circ}$ is the only case in which the rms prediction is larger than the original flow value. This is related to the energy overshoot discussed in the previous paragraph.

Finally, this manuscript proves that the energy added via the predicted fluctuations also reduces the error in the instantaneous flow field prediction. This is particularly true for the architecture with $d_{\text{model}}=256$. The temporal mean of the energy prediction error is reduced on all angles from more than the $37\%$ to less than the $12\%$. Moreover, the KL divergence between the velocity distribution reconstructed by POD and the one of the original field is consistently larger than ${\cal{{D_{KL}}}}=0.2$, but the closure reduces it to less than ${\cal{{D_{KL}}}}=0.026$.

\backsection[Acknowledgements]{The authors acknowledge the Barcelona Supercomputing Center for the usage of MareNostrum 5 during the development of this manuscript. The authors also acknowledge the insightful conversations with Fermín Mallor during the development of the initial ideas giving place to this work.}

\backsection[Funding]{The research leading to this work has been partially funded by the project TIFON with reference PLEC2023-010251/ AEI/10.13039/501100011033. B. Eiximeno’s work was funded by a contract from the Subprograma de Ayudas Predoctorales given by the Ministerio de Ciencia e Innovación (PRE2021-096927). Oriol Lehmkuhl has been partially supported by a Ramon y Cajal postdoctoral contract (Ref: RYC2018-025949-I). The authors acknowledge the support of Departament de Recerca i Universitats de la Generalitat de Catalunya to the Research Group Large-scale Computational Fluid Dynamics (Code: 2021 SGR 00902) and the Turbulence and Aerodynamics Research Group (Code: 2021 SGR 01051). We also acknowledge the Barcelona Supercomputing Center for awarding us access to the MareNostrum IV machine based in Barcelona, Spain.

Marcial Sanchis-Agudo and Ricardo Vinuesa would like to acknowledge the support from Marie Sklodowska-Curie Actions project MODELAIR, funded by the European Commission under the Horizon Europe program through grant agreement number 101072559. }

\backsection[Declaration of interests]{The authors report no conflict of interest.}

\backsection[Data availability statement]{The data that support the findings of this study is available upon request. See JFM's \href{https://www.cambridge.org/core/journals/journal-of-fluid-mechanics/information/journal-policies/research-transparency}{research transparency policy} for more information}

\backsection[Author ORCIDs]{B. Eiximeno, https://orcid.org/0000-0001-7018-6371; M. Sanchis-Agudo, https://orcid.org/0009-0000-1194-7900; A. Miró, https://orcid.org/0000-0002-2772-6050; I. Rodriguez, https://orcid.org/0000-0002-3749-277X; R. Vinuesa, https://orcid.org/0000-0001-6570-5499; O. Lehmkuhl, https://orcid.org/0000-0002-2670-1871}

\backsection[Author contributions]{\textbf{B. E.:} Writing – review \& editing, Writing – original draft, Visualization, Validation, Software, Methodology, Investigation, Formal analysis, Data curation, Conceptualization. \textbf{M. S-A.:} Writing – review \& editing, Writing – original draft, Software, Methodology, Investigation, Conceptualization. \textbf{A. M.:} Writing – review \& editing, Supervision, Software. \textbf{I. R.:} Writing – review \& editing, Supervision, Methodology. \textbf{R. V:} Writing – review \& editing, Supervision, Resources, Project administration, Funding acquisition. \textbf{O. L.:} Writing – review \& editing, Validation, Supervision, Software, Resources, Project administration, Methodology, Investigation, Funding acquisition.
}

\appendix
\section{Accuracy of the closure on the training angles}
\label{appA}
\begin{figure}
    \centering
    \subfloat[]{\includegraphics[width=\textwidth]{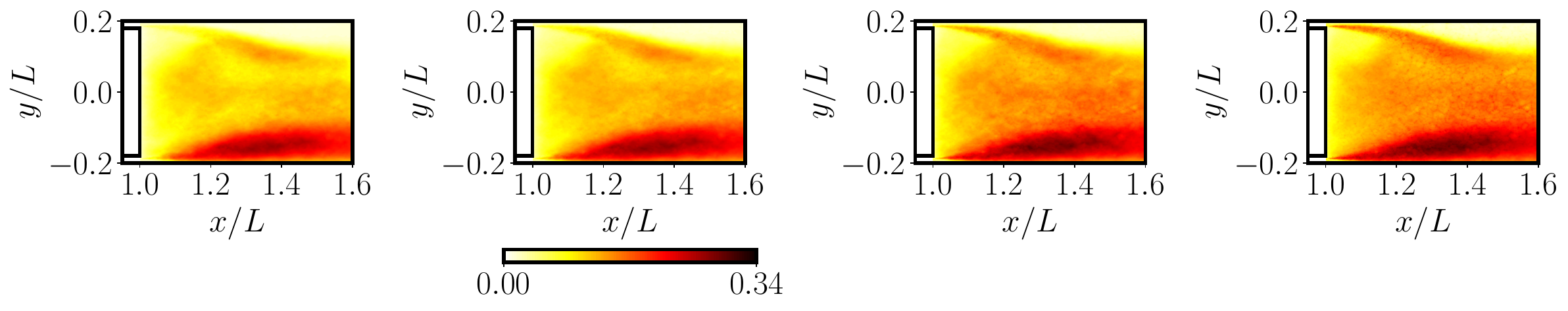}\label{fig:contour_5}}\\
    \subfloat[]{\includegraphics[width=\textwidth]{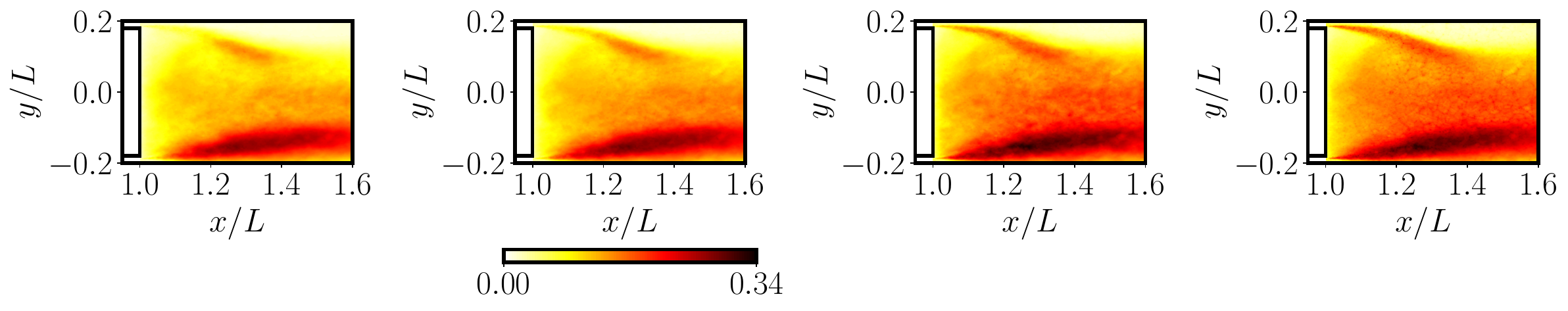}\label{fig:contour_10}}\\
    \subfloat[]{\includegraphics[width=\textwidth]{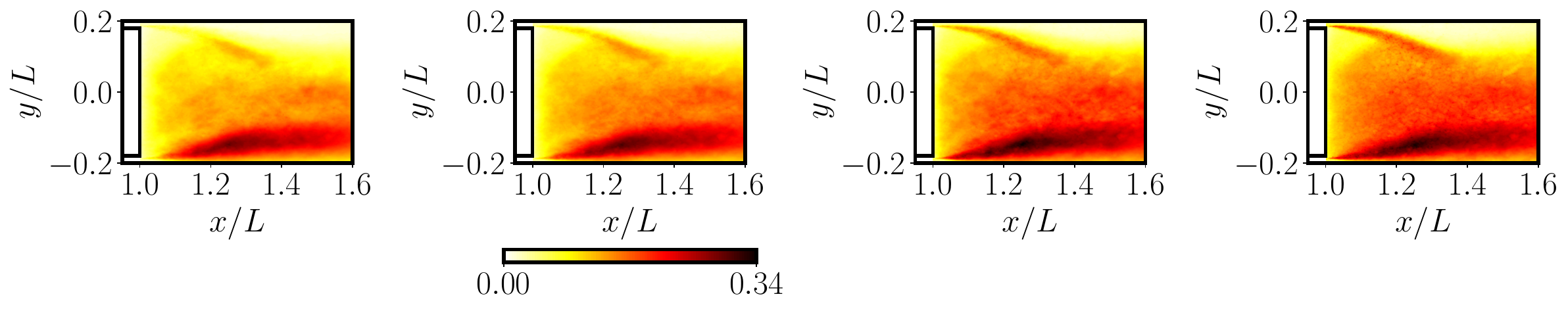}\label{fig:contour_12_5}}
   \caption{Root mean square of the streamwise velocity fluctuations for the cases at $\delta=5^{\circ}$ (a), $\delta=10^{\circ}$ (b) and $\delta=12.5^{\circ}$ (c). From left to right, the pictures represent: reconstruction from the common POD basis, reconstruction with the closure model with an attention size of $d_{\text{model}}=64$, reconstruction with the closure model with an attention size of $d_{\text{model}}=256$ and the original flow.}
   \label{fig:contour_architectures_train}
\end{figure}

\begin{figure}
    \centering
    \subfloat[]{\includegraphics[width=0.32\textwidth]{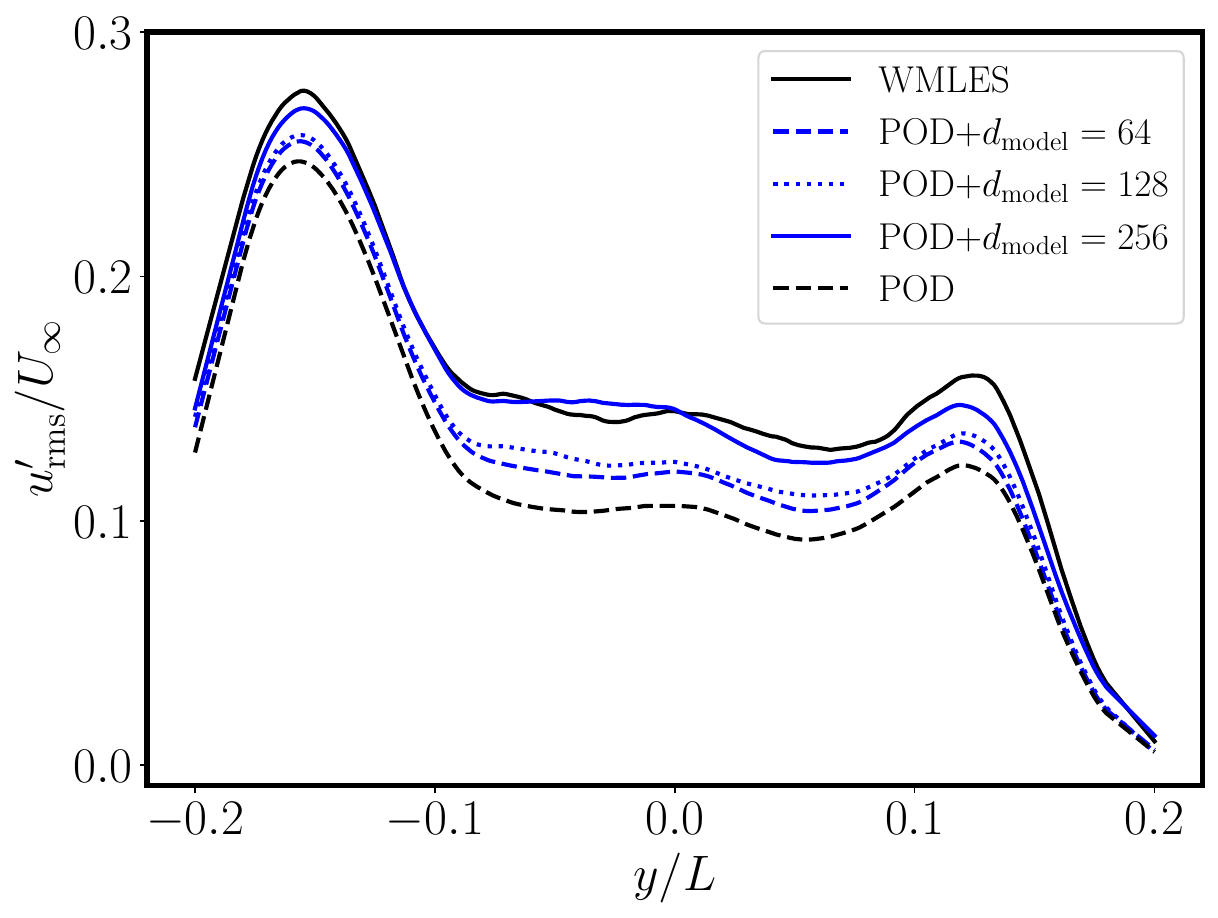}\label{fig:line_5}}
    \subfloat[]{\includegraphics[width=0.32\textwidth]{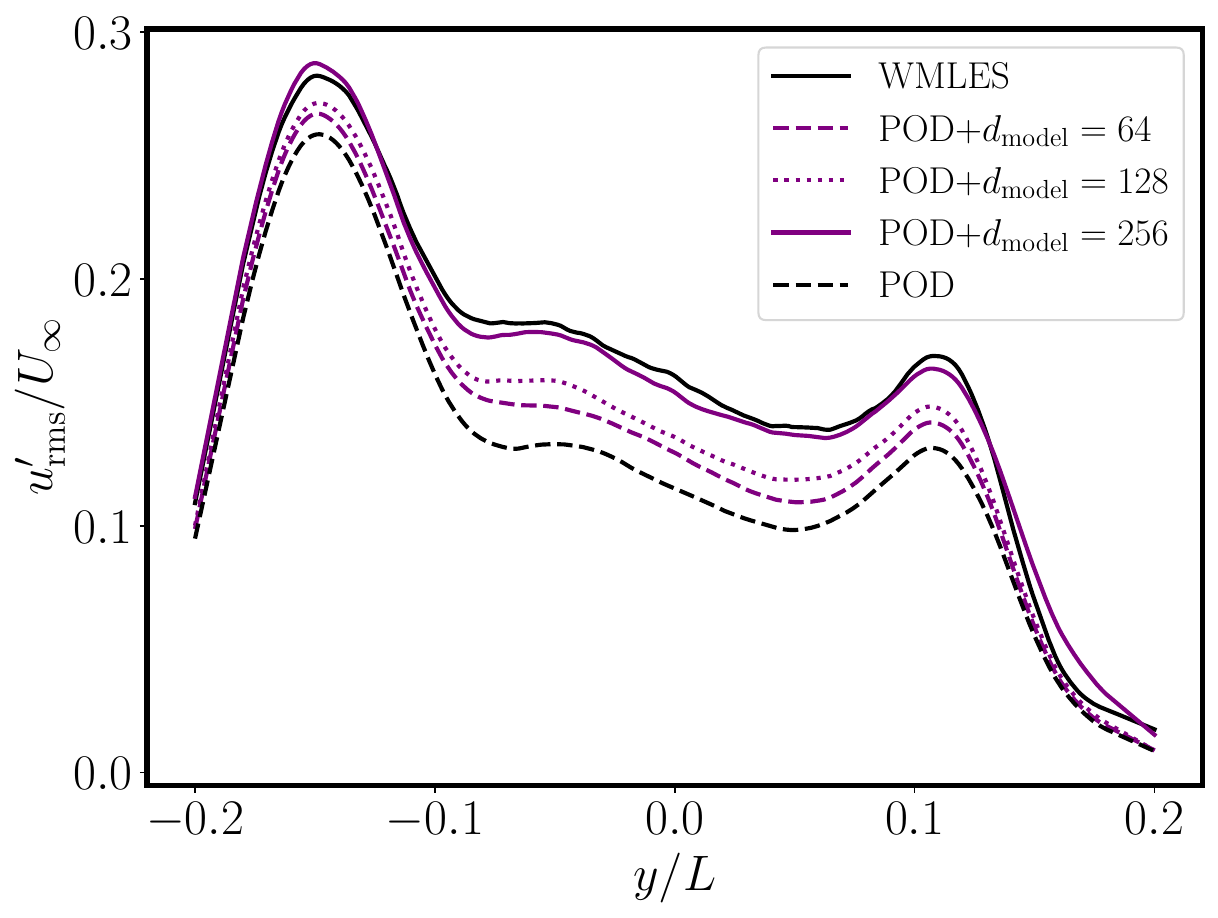}\label{fig:line_10}}
    \subfloat[]{\includegraphics[width=0.32\textwidth]{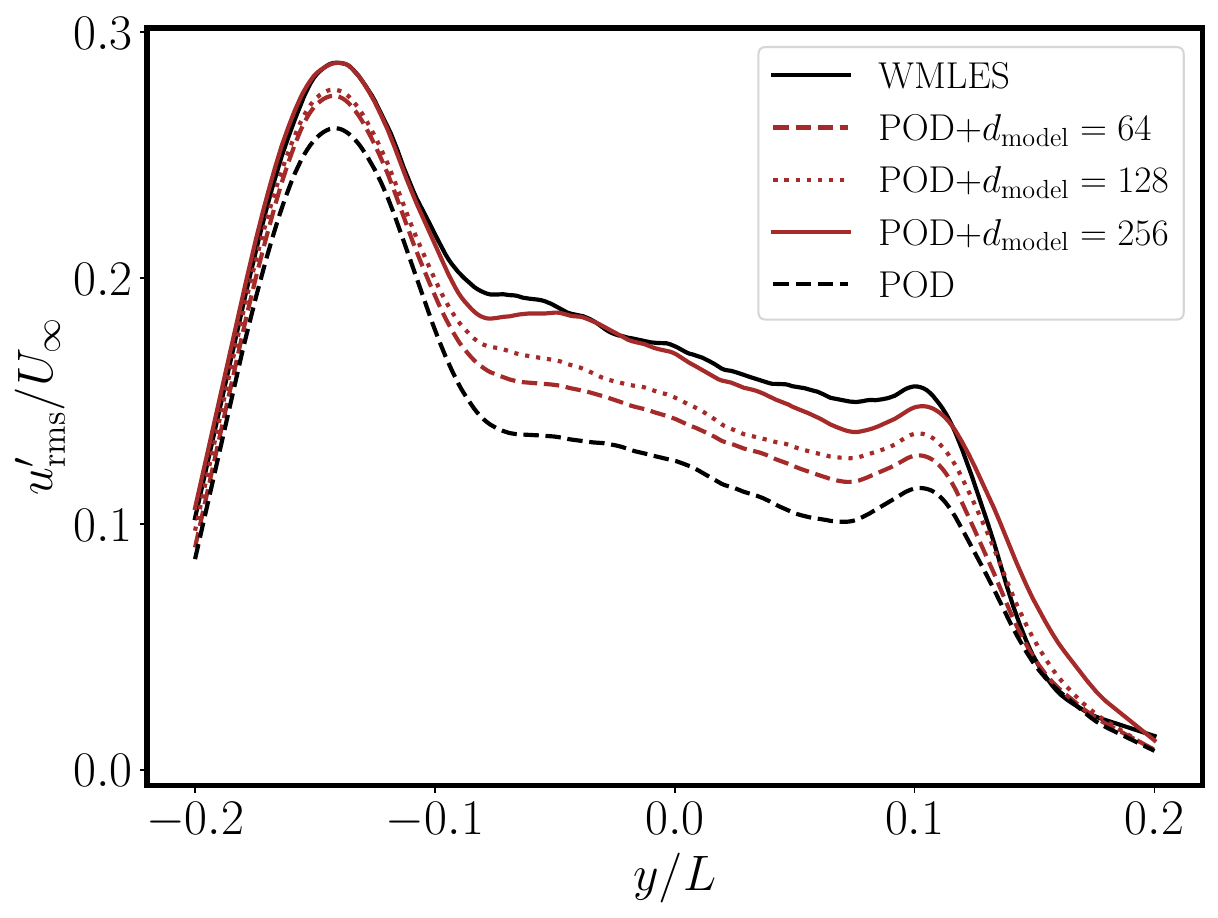}\label{fig:line_12_5}}
   \caption{Root mean square of the streamwise velocity fluctuations for the cases at $\delta=5^{\circ}$ (a), $\delta=10^{\circ}$ (b) and $\delta=12.5^{\circ}$ (c) at $x/L=1.3$.}
   \label{fig:line_architectures_train}
\end{figure}

This appendix presents the comparison of the root mean square value of the streamwise velocity fluctuations for the angles of $\delta=[5^{\circ}, 10^{\circ}, 12.5^{\circ}]$. Those angles were also included in the common basis as the case of $\delta=2.5^{\circ}$. Moreover, the error of their reconstruction was also included in the dataset used for the training of the transformer. \autoref{fig:contour_architectures_train} complements \autoref{fig:contour_architectures} and \autoref{fig:line_architectures_train} complements \autoref{fig:line_architectures} as in those figures only the case at $\delta=2.5^{\circ}$ was used to illustrate the effect of the closure on the cases used during training.

\bibliographystyle{jfm}
\bibliography{references}

\end{document}